\documentclass[a4paper,fleqn,usenatbib]{mnras}

\usepackage[T1]{fontenc}
\usepackage{ae,aecompl}

\usepackage{graphicx}	
\usepackage{amsmath}	
\usepackage{amssymb}	
\usepackage{comment}
\usepackage{color}
\usepackage{hyperref}

\title[3D AMR simulations of a compact source in G2]{3D AMR hydrosimulations of a compact source scenario for the Galactic Centre cloud G2}

\author[A. Ballone et al.]{A. Ballone$^{1,2}$, M. Schartmann$^{3}$, A. Burkert$^{1,2}$, S. Gillessen$^{2}$, P. M. Plewa$^{2}$,\newauthor R. Genzel$^{2}$, O. Pfuhl$^{2}$, F. Eisenhauer$^{2}$, M. Habibi$^{2}$, T. Ott$^{2}$ and E. M. George$^{2}$
\\
$^{1}$University Observatory Munich, Scheinerstra{\ss}e 1, D-81679 M{\"u}nchen, Germany\\
$^{2}$Max-Planck-Institute for Extraterrestrial Physics, Postfach 1312, Giessenbachstra{\ss}e, D-85741 Garching, Germany\\
$^{3}$Centre for Astrophysics and Supercomputing, Swinburne University of Technology, Hawthorn, Victoria 3122, Australia}

\date{Accepted XXX. Received YYY; in original form ZZZ}

\pubyear{2015}

\begin{document}
\label{firstpage}
\pagerange{\pageref{firstpage}--\pageref{lastpage}}
\maketitle

\begin{abstract}
The nature of the gaseous and dusty cloud G2 in the Galactic Centre is still under debate. We present three-dimensional hydrodynamical adaptive mesh refinement (AMR) simulations of G2, modeled as an outflow from a ``compact source'' moving on the observed orbit. The construction of mock position-velocity (PV) diagrams enables a direct comparison with observations and allow us to conclude that the observational properties of the gaseous component of G2 could be matched by a massive ($\dot{M}_\mathrm{w}=5\times 10^{-7} \;M_{\odot} \mathrm{yr^{-1}}$) and slow ($50 \;\mathrm{km \;s^{-1}}$) outflow, as observed for T Tauri stars. In order for this to be true, only the material at larger ($>100 \;\mathrm{AU}$) distances from the source must be actually emitting, otherwise G2 would appear too compact compared to the observed PV diagrams. On the other hand, the presence of a central dusty source might be able to explain the compactness of G2's dust component. In the present scenario, 5-10 years after pericentre the compact source should decouple from the previously ejected material, due to the hydrodynamic interaction of the latter with the surrounding hot and dense atmosphere. In this case, a new outflow should form, ahead of the previous one, which would be the smoking gun evidence for an outflow scenario.
\end{abstract}




\section{Introduction}\label{intro}

Since the date of its discovery, the nature of the little gaseous and dusty cloud G2 discovered by \citet{Gillessen12} has remained enigmatic. The Br$\gamma$, Pa$\alpha$ and HeI recombination lines detected with the integral field spectrographs SINFONI\footnote[1]{http://www.eso.org/sci/facilities/paranal/instruments/sinfoni/} at the VLT \citep{Gillessen12, Gillessen13a, Gillessen13b, Pfuhl15,Valencia-S15} and OSIRIS\footnote[2]{http://www2.keck.hawaii.edu/inst/osiris/} at the Keck telescope \citep{Phifer13} show a gas component extending both in size and velocity and following a high eccentricity Keplerian orbit \citep[see the position-velocity diagrams in][]{Gillessen13a,Gillessen13b,Pfuhl15}. \citet{Gillessen13b} and \citet{Pfuhl15} have also reported the detection of a blue-shifted component, simultaneous with the red-shifted one, consistent with G2 passing pericentre as an extended object in March-April 2014. 
The detections in L' and M' bands in the NIR with NACO\footnote[3]{http://www.eso.org/sci/facilities/paranal/instruments/naco/} at the VLT \citep{Gillessen12, Gillessen13a} and with NIRC2\footnote[4]{http://www2.keck.hawaii.edu/inst/nirc2/} at the Keck telescope \citep{Phifer13, Witzel14} suggest that G2 has an unresolved dust component at a temperature of roughly 550 K\footnote[5]{The unresolved nature of the dust component shows that the latter is more compact than the gaseous one. As a matter of fact, the large point spread function (PSF) of NACO and NIRC2 makes the size of the dust emission only marginally in discrepancy with the sizes inferred from the Br$\gamma$ emission. As shown by \citet{Witzel14}, the PSF in L' is also bigger than the tidal radius of a 2 $M_{\odot}$ star, hence any dusty material might still be considerably extended and unbound from a possible central object.}.

Several observational programs are currently monitoring its evolution\footnote[6]{https://wiki.mpe.mpg.de/gascloud/FrontPage}, also focusing on the interaction of this object with the extreme gravitational field of the $4.31\times10^6 \; M_{\odot}$ supermassive black hole (SMBH) centred on SgrA* \citep{Ghez08, Gillessen09} and with the hot and dense plasma accreting onto it. For example, increased emission in X-ray \citep{Gillessen12} and radio \citep{Narayan12,Sadowski13b,Sadowski13a,Crumley13,Abarca14} have been predicted by some models of the interaction of G2 with the outer accretion flow, but no consistent back reaction from either the accretion flow or SgrA* have been detected so far \citep{Haggard14, Chandler14, Bower15, Borkar16}. G2's partial or total disruption might also affect the accretion rate onto SgrA* or affect the statistics and properties of flares from SgrA*: \citet{Ponti15} showed that there has been an increase in the rate of X-ray bright flares since summer 2014, that might have been induced by G2's pericentre passage. However, this result is still under debate \citep[][]{Mossoux16} and further monitoring of SgrA* is needed to draw any strong conclusion. Finally, \citet{Plewa17} have recently presented SINFONI and NACO observations of G2 in 2015 and 2016. In these, G2 appears to have passed pericentre, keeping on following more or less the same predicted orbit.

These observations are performed with the most up-to-date instruments, pushing them to the limits of their capabilities; nonetheless, given the very small scales, it is still hard to evaluate the importance of the different physical processes in play. Trying to theoretically model the origin and fate of G2 has hence turned out to be challenging, but in the last two years several studies have shed light on this peculiar object.

The theoretical picture is presenting a dichotomy: G2 could be either a clump of diffuse gas and dust plunging into SgrA* or the outflow from a central source (possibly a young star) on a high eccentricity orbit around the SMBH.

The first scenario has been originally proposed by \citet{Gillessen12}. In this context, the gas is at a temperature $T\approx 10^4 \;\mathrm{K}$ and it is fully ionized by the ultraviolet (UV) radiation field produced by the nearby young and massive stars. Under the assumption of case-B recombination and of a homogeneous sphere of radius $R_c\approx 1.9\times10^{15} \;\mathrm{cm}$, these authors derived a mass of $M_{\mathrm{G2}}\approx 1.7 \times 10^{28}\;\mathrm{g}\approx 3$ Earth masses and a uniform density of $\rho_c\approx6.1\times10^{-19} \mathrm{\;g\;cm^{-3}}$. Several studies have been carried out for this scenario, focusing on the evolution of G2 and on its interaction with SgrA*'s accretion flow \citep{Burkert12,Schartmann12, Anninos12, Shcherbakov14,Schartmann15}. G2 seems to be followed by a larger component (named G2t or ``the tail'') following G2 on a similar orbit \citep{Gillessen13a, Pfuhl15, Plewa17} and \citet{Pfuhl15} have shown that G2 could be connected to the previously discovered gas/dust cloud G1 \citep{Clenet04a,Clenet04b,Clenet05,Ghez05}, whose orbit can be matched by a G2-like orbit after a drag force is applied to it (\citealp{Pfuhl15,McCourt16,Madigan17}; but see \citealp{Plewa17} for a different finding). This observational finding suggests that G2 is actually part of a much larger streamer. The idea of a gas streamer has been already proposed by \citet{Guillochon14}, where the streamer could be produced by tidal stripping of the outer envelope of a late-type giant star, in a close encounter of such a star with the central SMBH. Another possibility for the origin of G2, if G2 is not linked to a central object, is clump formation through the non-linear thin shell instability in colliding winds of the outer O/WR stars \citep{Calderon16}.

The second scenario involves a connection with a central source on G2's orbit. G2's Br$\gamma$ emission could either result from the gas lost by a photoevaporating disk \citep{Murray-Clay12, Miralda-Escude12} or by a photoevaporating starless (proto-)planet, tidally captured by the SMBH \citep{Mapelli15, Trani16} or produced by the interaction between an outflow from a low-mass star and the hot accretion flow \citep{Scoville13,Ballone13,DeColle14,Zajacek14, Zajacek16,Zajacek17} or a nova outburst \citep{Meyer12}. \citet{Valencia-S15} tried to fully explain the Br$\gamma$ line-width with a combination of an accretion stream and a disk wind close to a low-mass star. However, this is not in agreement with the PV diagrams obtained by \citet{Gillessen13b} and \citet{Pfuhl15}, showing a spatially resolved velocity gradient consistent with tidal stretching. Finally, given the unresolved and constant-luminosity L'-band emission, \citet{Witzel14} hypothesized that G2 is a binary star merger \citep[see also][]{Prodan15,Stephan16}, forming a new low-mass ($<2 M_{\odot}$) star and heating the dust component from inside. In a recent work, \citet{Ballone16} showed that a relatively fast and massive outflow might also be able to reproduce both G2 and G2t at the same time, however neglecting the possible connection with the cloud G1. Differently from the present more quantitative study, focusing on reproducing only G2, the one in \citet{Ballone16} is rather meant to be a proof of concept. As already mentioned, observations keep on hinting that G2 and G2t are closely related, but their connection is not fully established, yet. This led us to test both scenarios; the link and differences between the two studies are discussed in section \ref{seccomp}.

In this paper we focus on G2 only and present 3D simulations of an outflow scenario. Compared to the 2D simulations in \citet{Ballone13}, 3D simulations represent the geometry of the problem in a more realistic way and allow a much stricter comparison with the observations. Unfortunately, the high resolution used with the 2D simulations in \citet{Ballone13} cannot be reached in this 3D study. So, the current simulations should be thought as complementary to the 2D ones presented in \citet{Ballone13}, rather than simple upgrades of them.

In Section \ref{secsetup} we describe the setup of our simulations. The results are presented in Section \ref{secresults}, where we compare them to the observations and we study the effect of the outflow parameters. Section \ref{secdisc} is dedicated to a more careful discussion of the ionization of the gas and the related uncertainties and of the numerical limitations. We also compare our study with previous ones and present the advantages and disadvantages of such a scenario. Summary and final remarks can be found in Section \ref{secsum}.

\begin{table*}
\caption{Parameters of the simulated 3D AMR models.}
\label{parsimul}
\begin{tabular}{l l l l l l}
\hline
 & $\dot{M}\mathrm{_w} \mathrm{(M_{\odot} \;yr^{-1})}$ & $v\mathrm{_w} \mathrm{(km/s)}$ & max resolution & coordinates & domain size (x $\times$ y $\times$ z / R $\times$ z) ($10^{16}$ cm) \\
\hline
standard model & $\mathrm{5 \times 10^{-7}}$ & 50 & $1.25\times10^{14}$ cm & 3D cartesian & $[-26.4:1.2]\times[-3.6:4.8]\times[-2.4:2.4]$ \\
 & & & 8.3 AU & (AMR) & \\
HV3D & $\mathrm{5 \times 10^{-7}}$ & 250 & $1.25\times10^{14}$ cm & 3D cartesian & $[-28.8:2.4]\times[-3.6:7.2]\times[-4.8:4.8]$\\
 & & & 8.3 AU & (AMR) & \\
LMDOT3D & $\mathrm{10^{-7}}$ & 50 & $1.25\times10^{14}$ cm & 3D cartesian & $[-26.4:1.2]\times[-3.6:4.8]\times[-2.4:2.4]$\\
 & & & 8.3 AU & (AMR) & \\
HMDOT3D & $\mathrm{2.5 \times 10^{-6}}$ & 50 & $1.25\times10^{14}$ cm & 3D cartesian & $[-26.4:1.2]\times[-3.6:4.8]\times[-2.4:2.4]$\\
 & & & 8.3 AU & (AMR) & \\
stLOWRES & $\mathrm{5 \times 10^{-7}}$ & 50 & $2.5\times10^{14}$ cm & 3D cartesian & $[-26.4:1.2]\times[-3.6:4.8]\times[-2.4:2.4]$ \\
 & & & 16.6 AU & (AMR) & \\
st2D & $\mathrm{5 \times 10^{-7}}$ & 50 & $1.25\times10^{14}$ cm & 2D cylindrical& $[0.0:1.8]\times[-28.8:-3.0]$ \\
 & & & 8.8 AU & (fixed grid) & \\
\hline
\end{tabular}
\end{table*}

\begin{figure*}
\begin{center}
\includegraphics[scale=0.45]{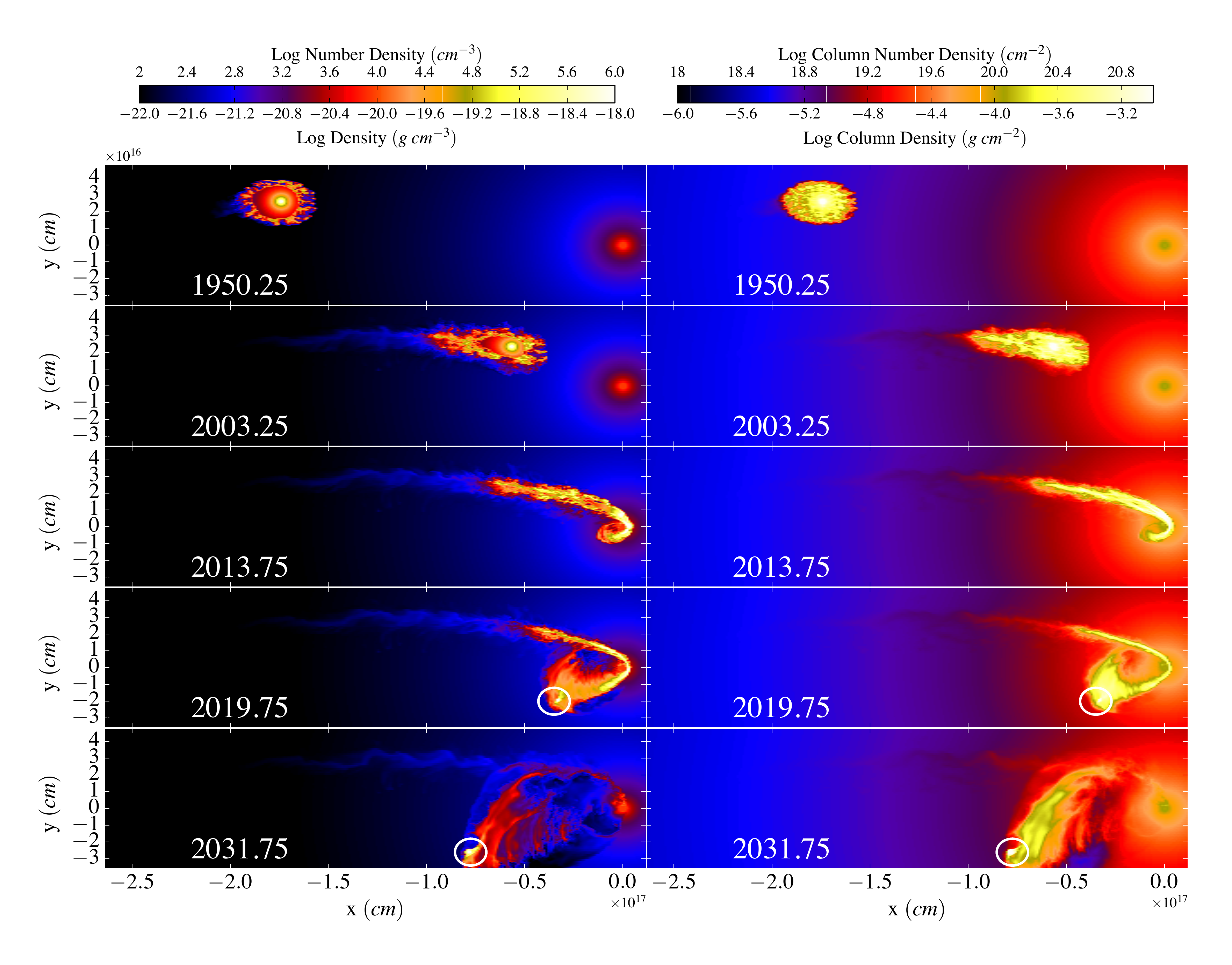}
\caption{Density maps for the standard model. Left panels show the density distribution in a slice at z=0. The right panels show the column density, i.e. the integral of the density along the z direction. The white circles show the outflow reforming after pericentre.
}\label{denmap}
\end{center}
\end{figure*}

\section{Simulation setup}\label{secsetup}

The simulations presented in this paper were run with the Eulerian code PLUTO \citep{Mignone07,Mignone12}. Performing these three dimensional simulations with a uniform grid is computationally not feasible \citep[see discussion in][]{Ballone13}, so we adopted the adaptive mesh refinement (AMR) strategy, implemented in the code through the CHOMBO\footnote[1]{https://seesar.lbl.gov/anag/chombo/} library. For the refinement criterion, we chose the standard one in PLUTO, based on the second derivative error norm, and we applied it to the density. The criterion has been widely tested and it is able to resolve most of G2's material at the highest resolutions. The computational domain is Cartesian (with the exception of one test run in 2D cylindrical coordinates, see Table \ref{parsimul}). 
A two-shock Riemann solver \citep{Mignone12} has been chosen for the solution of the hydrodynamic equations. 

The outflow is modeled in a ``mechanical'' way as in \citet{Ballone13,Ballone16} and \citet{DeColle14}, where the velocity is set to the constant wind value $v_\mathrm{w}$ and the density $\rho_\mathrm{w}$ is set to satisfy  

\begin{equation}\label{conteq}
\dot{M}_\mathrm{w}=4\pi r_\mathrm{w}^2\rho_\mathrm{w}v_\mathrm{w}.
\end{equation}

In order to reach a reasonable sampling of the input region, i.e. a good isotropy of the outflow, the input region's radius $r_\mathrm{w}$ is varying, in time, proportional to the theoretical stagnation radius $R_{\mathrm{out}}$ (see Eq. \ref{stag}), with minimum and maximum allowed values equal to $2.10\times10^{14}\;\mathrm{cm}$ and $1.05\times10^{15}\;\mathrm{cm}$, respectively.
The temperature of the injected material is set to $T_\mathrm{w}=10^4 \; \mathrm{K}$ and an adiabatic index $\Gamma=1$ has been assumed \citep[see discussion in][]{Ballone13}.

Compared to \citet{Ballone13}, the source's orbit is now a proper elliptical orbit and it has been updated to the one derived by \citet{Gillessen13b} through Br$\gamma$ observations. The orbit has been previously integrated with a leapfrog method and the source's positions and velocities are interpolated from the stored ones using a 1st order Newton polynomial formula.
 
The hot atmosphere is modeled following the density and temperature distribution used by several authors \citep[see][]{Burkert12,Schartmann12,Anninos12,Ballone13,DeColle14,Ballone16}, i.e., 

\begin{equation}\label{denatm}
n\mathrm{_{at}}\simeq5.60\times10^3 \left(\frac{1}{d_\mathrm{{BH,peri}}}\right)\;\mathrm{cm^{-3}},
\end{equation}
\begin{equation}\label{tematm}
T\mathrm{_{at}}\simeq7.12\times 10^8\left(\frac{1}{d_\mathrm{{BH,peri}}}\right)\mathrm{\; K},
\end{equation}

where $d_{\mathrm{BH,peri}}$ is the distance from SgrA* in units of the pericentre distance, i.e. $3\times10^{15}\; \mathrm{cm}$.

This is a very idealized model and, given the uncertainties in the actual distribution of the accretion flow around SgrA*, we still decided to keep it as idealized as possible, to be able to better understand 0th-order hydrodynamical effects on G2. This would be difficult when doing more sophisticated modeling. As in \citet{Schartmann12,Schartmann15} and \citet{Ballone13,Ballone16}, we reset the atmosphere with the help of a passive tracer. In order to reproduce the outer shock propagating in the atmosphere, \citet{DeColle14} did not apply the same recipe for two of their simulations. However, in these cases, the development of convective bubbles all around the SMBH region is apparent. This artifact is avoided in our approach.
Finally, the SMBH's gravitational field has been modeled as a Newtonian point source with mass $M_\mathrm{{BH}} = 4.31 \times 10^6 M_{\odot}$ \citep{Gillessen09} at $x, y, z= 0$. We refer to \citet{Ballone13} for further discussions and details about the modeling and the assumptions.

As in \citet{Ballone13,Ballone16} and differently than in \citet{DeColle14}, we decided to start the simulation (and the outflow) at apocentre. As already pointed out in \citet{Ballone13}, this choice is somehow arbitrary. However, if the source of G2 had been scattered via multiple encounters \citep{Murray-Clay12} from the clockwise rotating disk of young stars \citep{Paumard06, Bartko09}, any pre-existing gas envelope would have been tidally torn apart. 

A list of the simulations discussed in the present paper can be found in Table \ref{parsimul}.

\section{Results}\label{secresults}

The purpose of this section is to present the evolution of our new 3D simulations and their comparison to observations, in which we will focus on the new and accurate construction of mock PV diagrams and on the interpretation of the time dependence of the total Brackett-$\gamma$ luminosity.

\subsection{The standard model}\label{secstand}

We adopt a mass loss rate of $\dot{M}\mathrm{_w=5\times 10^{-7} M_{\odot} \;yr^{-1}}$ and a wind velocity of $v\mathrm{_w = 50 \;km\, s^{-1}}$ for our standard model.

As shown in Fig. \ref{denmap}, the evolution of the density distribution in this 3D simulation is very similar to the one of the 2D simulations in \citet{Ballone13} \citep[for an exhaustive discussion on the physics of these winds, we also refer the reader to][]{Christie16,Ballone16}. The outflow is free-flowing until its ram pressure reaches the pressure of the external hot and dense atmosphere. Hence, it is composed of an inner part, whose density scales as $1/r^2$ (due to the continuity equation), that is surrounded by the part of the outflow that gets shocked by the impact with the atmosphere. This shocked material is highly Rayleigh-Taylor unstable. At the beginning, the outflowing material is still in a quasi-spherical configuration, since the isotropic thermal pressure of the atmosphere is still dominant compared to the anisotropic ram pressure. At later times, the free-wind region shrinks due to the increasing thermal pressure, the ram pressure makes it asymmetric and the stripped shocked material is forming a small tail trailing the source. Overall, though more filamentary, the distribution of the outflowing gas is on large scales very similar to the one in the ``diffuse cloud'' simulations of \citet{Schartmann12}, \citet{Anninos12} and \citet{Schartmann15}, particularly right before and after pericentre, when the material is first compressed into a thin filament by the tidal force from the SMBH and then expands, strongly increasing its cross section.

Due to the asymmetry of the free-wind region and the formation of the small tail of stripped material, at the time of the observations, the central source is always in the leading part of G2. The immediate implication is that the photocentre of the emission will never be on top of the source.

As expected, the simulation also shows that the emitting source becomes, at a certain point, distinguishable from the rest of G2. This might happen already around year 2019-2020, when the source creates a second peak in the density distribution (see circles in the lowermost panels of Fig. \ref{denmap}). This is a clear difference compared to the diffuse cloud simulations and the decoupling between the source and the previously emitted gas, after pericentre, could eventually be the smoking gun to understand the nature of G2.

\begin{figure*}
\begin{center}
\includegraphics[scale=0.18]{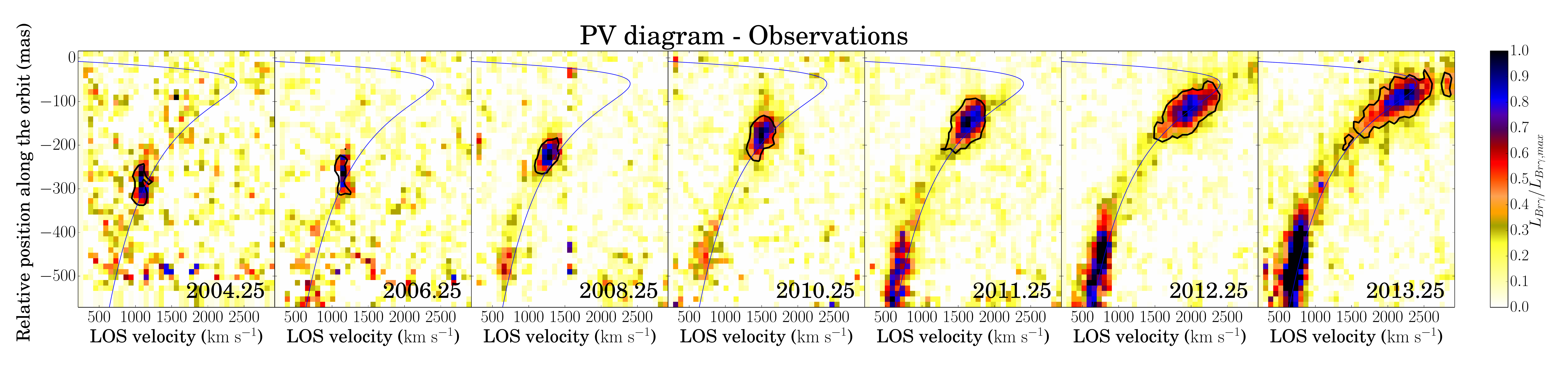}
\includegraphics[scale=0.18]{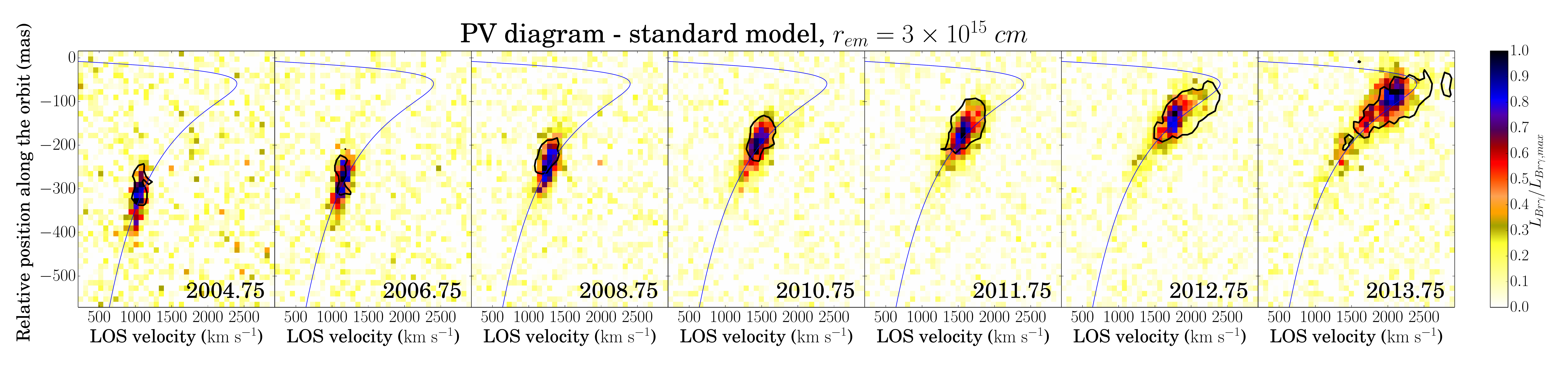}
\caption{Position-Velocity diagrams for the standard model. The upper panel shows the observations, while the lower one shows the case of $r_\mathrm{{em}}= 3\times10^{15} \; \mathrm{cm}$. The black contours show the position and extent of the observed G2.
}\label{pvstandard}
\end{center}
\end{figure*}

\begin{figure*}
\begin{center}
\includegraphics[scale=0.52]{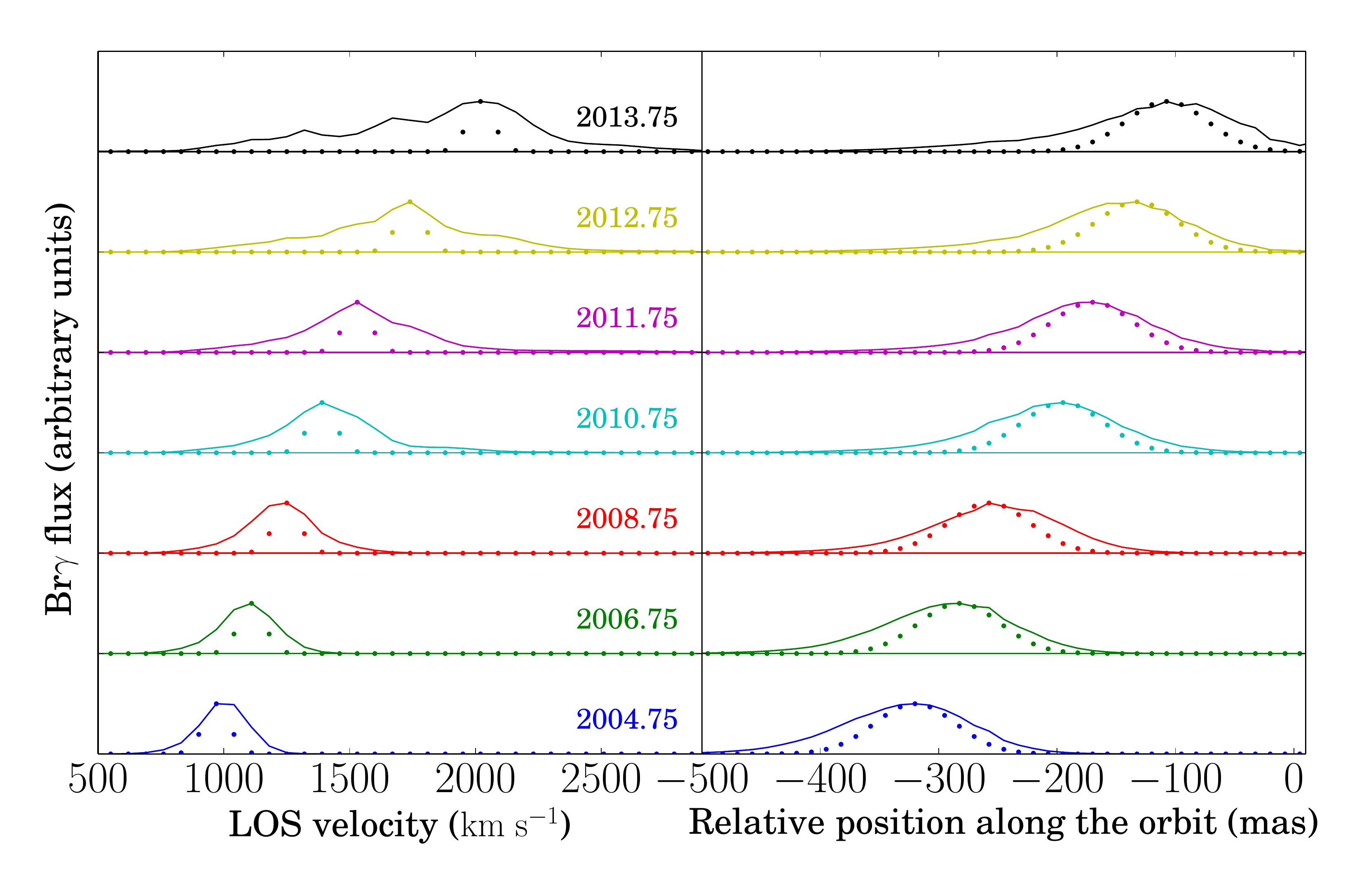}
\caption{Line width (left) and size along the orbit (right) evolution for the standard model, in the case of $r_\mathrm{{em}}= 3\times10^{15} \; \mathrm{cm}$ (solid lines). The dots represent a gaussian distribution centred on the peak of the emission, with FWHM equal to 120 $\mathrm{km \;s^{-1}}$ in velocity and 81 mas in the size, corresponding to the spectral and spatial point spread function of SINFONI.
}\label{linesplot}
\end{center}
\end{figure*}

\subsection{Matching the PV diagrams}

Compared to \citet{Ballone13}, the 3D simulation now allows us to construct realistic PV diagrams, like the ones already presented in \citet{Schartmann15}. To do this, we first project every cell in our computational domain onto the sky plane, according to the last orbital elements derived by \citet{Gillessen13b} for the Br$\gamma$ observations. This is done through a transformation from $(x,y,z,v_x,v_y,v_z)$ to $(ra,decl,losv)$ where $ra$, $decl$ and $losv$ are the right ascension, the declination and the line of sight (l.o.s.) velocity, respectively. We can, from this, create a 3D histogram of the Br$\gamma$ luminosity, with bin size equal to 12.5 mas for $ra$ and $decl$ and 69.6 $\mathrm{km\;s^{-1}}$ for $losv$. These values correspond to the size of the 3D pixels (voxels) in a SINFONI data cube. We then apply a smoothing in all directions with FWHM equal to 81 mas in right ascension and declination and to 120 $\mathrm{km\;s^{-1}}$  in l.o.s. velocity. These values correspond to the spatial point spread function (PSF) and spectral resolution. At this point, every cell is spatially projected onto the derived orbit, using it as a curved slit in the $(ra,decl)$ space \citep[a slit curved along G2's orbit has also been used for the construction of the observed PV diagrams; see][]{Gillessen13a}. The former operation reduces the triplet $(ra,decl,losv)$ to a couple $(pos,losv)$, where $pos$ is the position on the orbit, and creates a 2D position-velocity histogram. Given the uncertainties in the luminosity discussed in Sec. \ref{secion} and \ref{seclum}, every PV diagram is then scaled to its maximum. Noise is finally extracted from the observed PV diagrams and added to the simulated ones.

The luminosity is calculated using a functional form for the case B recombination Br$\gamma$ emissivity

\begin{equation}\label{lumform}
j\mathrm{_{Br\gamma}}= 3.44\times10^{-27}  (T/10^4\mathrm{\;K})^{-1.09}n_\mathrm{i}n\mathrm{_e \;erg\;s^{-1}\;cm^{3}},
\end{equation}
(where $T$ is the wind material temperature and $n_\mathrm{i}$ and $n_\mathrm{e}$ are the ion and electron number densities), obtained by extrapolating the values given on page 73 in \citet{Osterbrock06} \citep[see also][]{Ferland80,Hamann99,Ballone13}.

In Sec. \ref{secion} we will show that the amount of emission coming from the free flowing part of the outflow is uncertain. For an outflow scenario, this is strongly dependent on the flux of ionizing photons reaching G2, which is not exactly constrained.
For this reason, we present here the effect of different contributions on the total Br$\gamma$ luminosity of the free flowing region. Namely, we calculate PV diagrams assuming that the latter is ionized and emits in Br$\gamma$ only up to a certain inner radius $r_\mathrm{{em}}$. $r_\mathrm{{em}}$ is hence a free parameter of our post-processing and we choose $r_\mathrm{{em}}=[3\times10^{14},10^{15},3\times10^{15}] \; \mathrm{cm}$. We also calculated PV diagrams for the shocked outflow material only (in the text we will denote this case with $r_\mathrm{{em}}=r_\mathrm{{shock}}$).

\begin{figure}
\begin{center}
\includegraphics[scale=0.14]{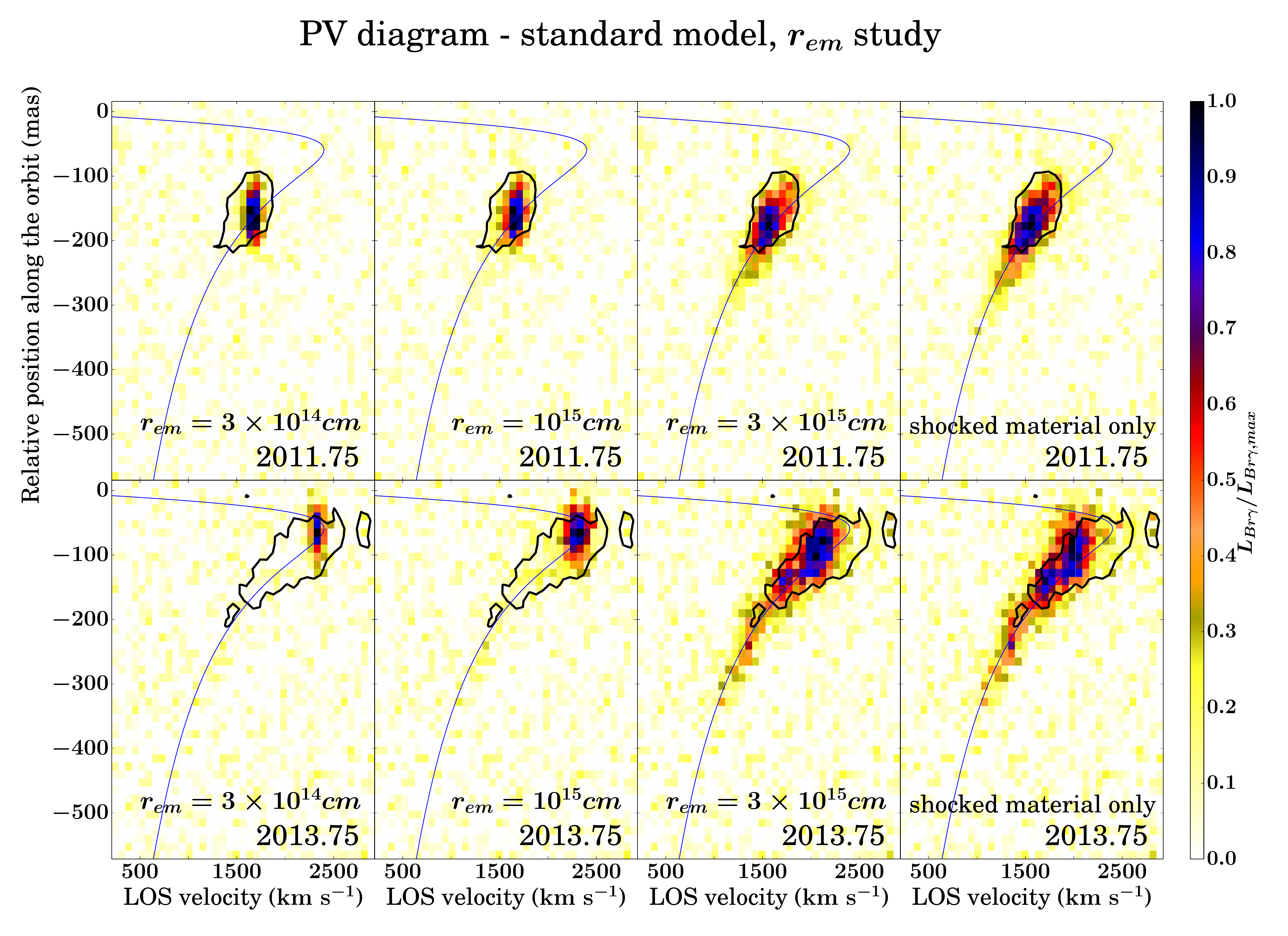}
\caption{Position-Velocity diagrams for the standard model. The different panels show the simulated PV diagrams for different assumptions on the inner emitting radius $r_\mathrm{{em}}$. For every panel, the luminosity per bin is scaled to the maximum one. The upper and lower panels are obtained for a simulation year of 2011.75 and 2013.75, respectively. The black contours show the position and extent of the observed G2.
}\label{pvrem}
\end{center}
\end{figure}

\begin{figure*}
\begin{center}
\includegraphics[scale=0.52]{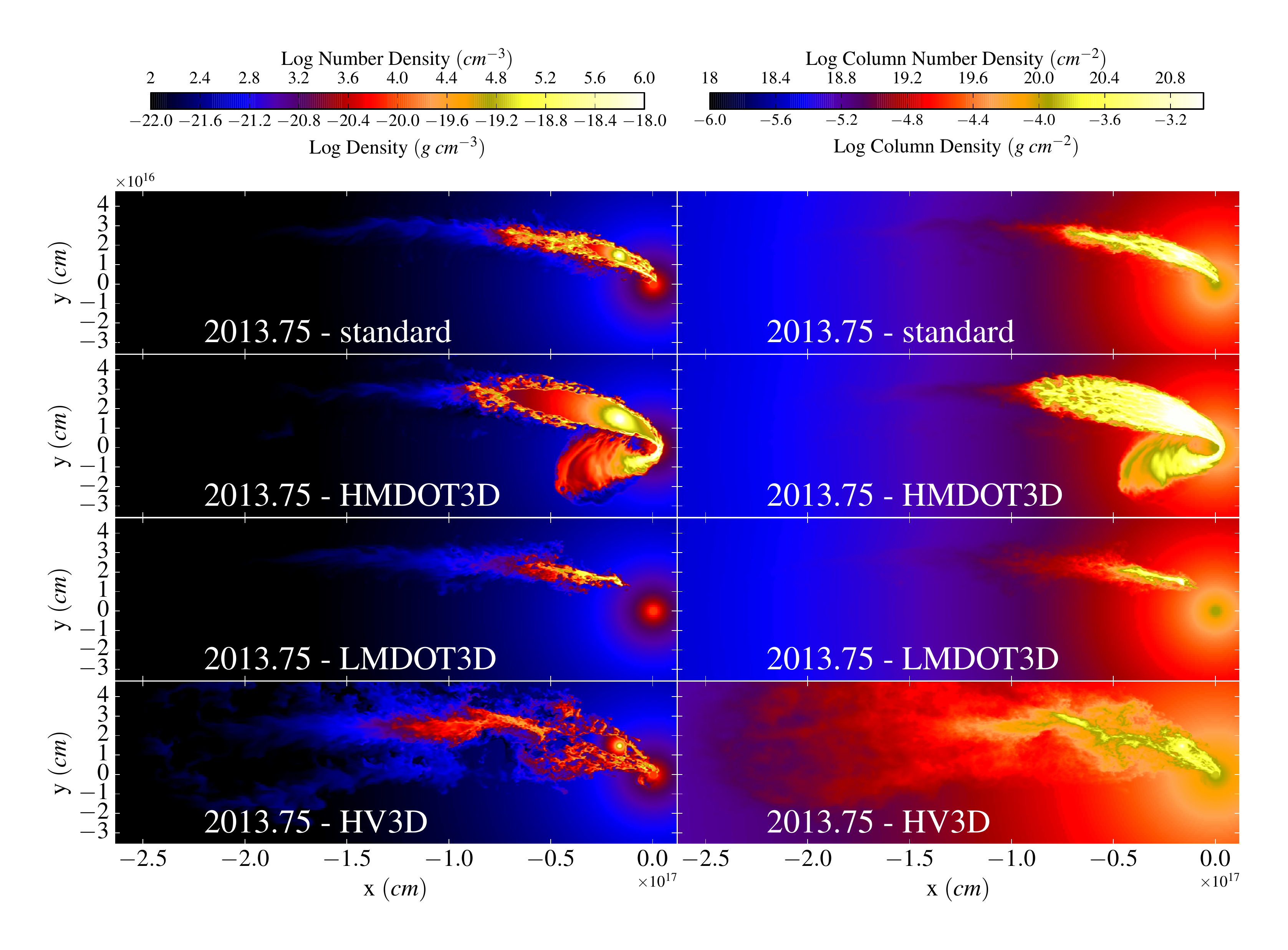}
\caption{Density maps for the simulations of our parameter study. Left panels show the density distribution in a slice at z=0. The right panels show the column density, i.e. the integral of the density along the z direction. 
}\label{denmappar}
\end{center}
\end{figure*}

The results are shown in Fig. \ref{pvstandard}, \ref{linesplot} and \ref{pvrem}. Due to the particularly dramatic evolution of the gas during and after pericentre, the already mentioned issues related to the luminosity discussed in Sec. \ref{secion} and \ref{seclum} are significantly effecting the reliability of our mock PV diagrams after 2014.5. For this reason, we restrict our comparison to the pre-pericentre part of the orbit and avoid making predictions for post-pericentre PV diagrams. As mentioned in Sec. \ref{secstand}, the orbital solution of the centre of the emission is never equal to that of the source; our comparison uses a time offset of roughly half a year between simulation and observations. The match is not perfect. In principle, a quantitative comparison between observed and simulated PV diagrams might eventually be used, through an iterative repetition of the simulation, to determine the orbit of the source that produces the perfect match. However, the high number of parameters of our models might not allow a strong constraint of the source's orbit and the high computational cost of these hydrodynamic simulations does not allow such a numerical experiment. Our purpose is rather to show which mass-loss rates and velocities an outflow should have to produce a reasonable result. This is already not trivial. Hence, we stick to a qualitative comparison and use a simple constant time offset. We must stress, thus, that this offset has no strong quantitative meaning.
Nonetheless, as visible in Fig. \ref{pvstandard} and \ref{linesplot}, our standard case is able to reproduce an increase in the line-width of the Br$\gamma$ emission, as in the case of the observations of G2. These figures also show that our standard model appears marginally resolved, even with some broadening of G2's size with its pericentre approach, as shown by \citet{Gillessen13b} and \citet{Pfuhl15}.

When looking at the $r_\mathrm{{em}}= r_\mathrm{{shock}}, 3\times10^{15} \;\mathrm{ cm}$ cases, the simulated material has a qualitatively comparable extent, even though it fails to reproduce the high velocity material that seems to overshoot the orbit derived from observations, just before the pericentre passage. This is again due to the fact that, close to pericentre, the outflow material is asymmetrically distributed with respect to the source, with most of the material in a trailing tail. When going to smaller values of $r_\mathrm{{em}}$, the emitting region moves to slightly higher velocities and positions on the orbit, but it becomes smaller and smaller in the PV diagrams. This is a direct consequence of the location and of the important impact of the free-wind region on the outflow emission. In fact, given Eq. \ref{conteq}, the emission measure EM$\propto\int \rho_\mathrm{w}dV\propto r^{-1}$ is diverging for small distances from the source. As a result, the more the inner part of the free-wind region is included, the more dominant the free-wind region, the smaller the emitting region visible in the PV diagrams. 

So, all in all, we conclude that a good match with the observations can be reached only if a tiny fraction of the free-wind region is actually emitting. This conclusion is general and can be also deduced from the parameter study in Sec. \ref{secparam}, where we show that G2 appears too small for every model, when $r_\mathrm{{em}} < 3\times 10 ^{15}\;\mathrm{cm}$. A probably better result could also be reached with a slightly different (more eccentric) orbital solution. In fact, uncertainties in the observations seem to give enough room for this possibility. However, testing it directly with simulations is beyond the scope of the present work.

\subsection{Parameter study}\label{secparam}

Following \citet{Ballone13}, we performed a parameter study, varying the mass-loss rate and the velocity of the outflow. We hence run models LMDOT3D and HMDOT3D with the same velocity as the standard model's one, but with a factor of 5 smaller and larger mass-loss rate, respectively. Concerning the velocity, we chose to run just the HV3D model, with wind velocity equal to $v_\mathrm{w}=250 \;\mathrm{km \;s^{-1}}=5\times v_\mathrm{{w,standard}}$. As already discussed in \citet{Ballone13}, given the isothermal equation of state, a temperature of $T=10^4 \; \mathrm{K}$ in the injected material brings the sound speed of the wind to $c_\mathrm{{s,w}}\approx 10\; \mathrm{km \;s^{-1}}$. As a consequence, for wind velocities too close to $c_\mathrm{{s,w}}$, the injected thermal and ram pressure become comparable, leading to too high mass loss rates and velocities. However, \citet{Ballone13} have already shown that a lower outflow velocity has the effect of reducing G2's size.

Fig. \ref{denmappar} shows the density maps for the three models of the parameter study. As already described in \citet{Ballone13} and \citet{Ballone16}, for LMDOT3D and HMDOT3D the outflow is too dense for the ram-pressure stripping to be efficient enough. Hence, the size of the outflow is mainly given by momentum equilibrium between the outflow and the external forces, namely the thermal and ram pressures of the atmosphere and the tidal force of the SMBH. This explains why LMDOT3D and HMDOT3D are respectively smaller and bigger than the standard model. In the HV3D case, the outflow is much less dense and the shocked material spreads out over a large volume. This enables the formation of a long cometary tail by efficient ram pressure stripping, as in the case of the model in \citet{Ballone16}.

\begin{figure}
\begin{center}
\includegraphics[scale=0.45]{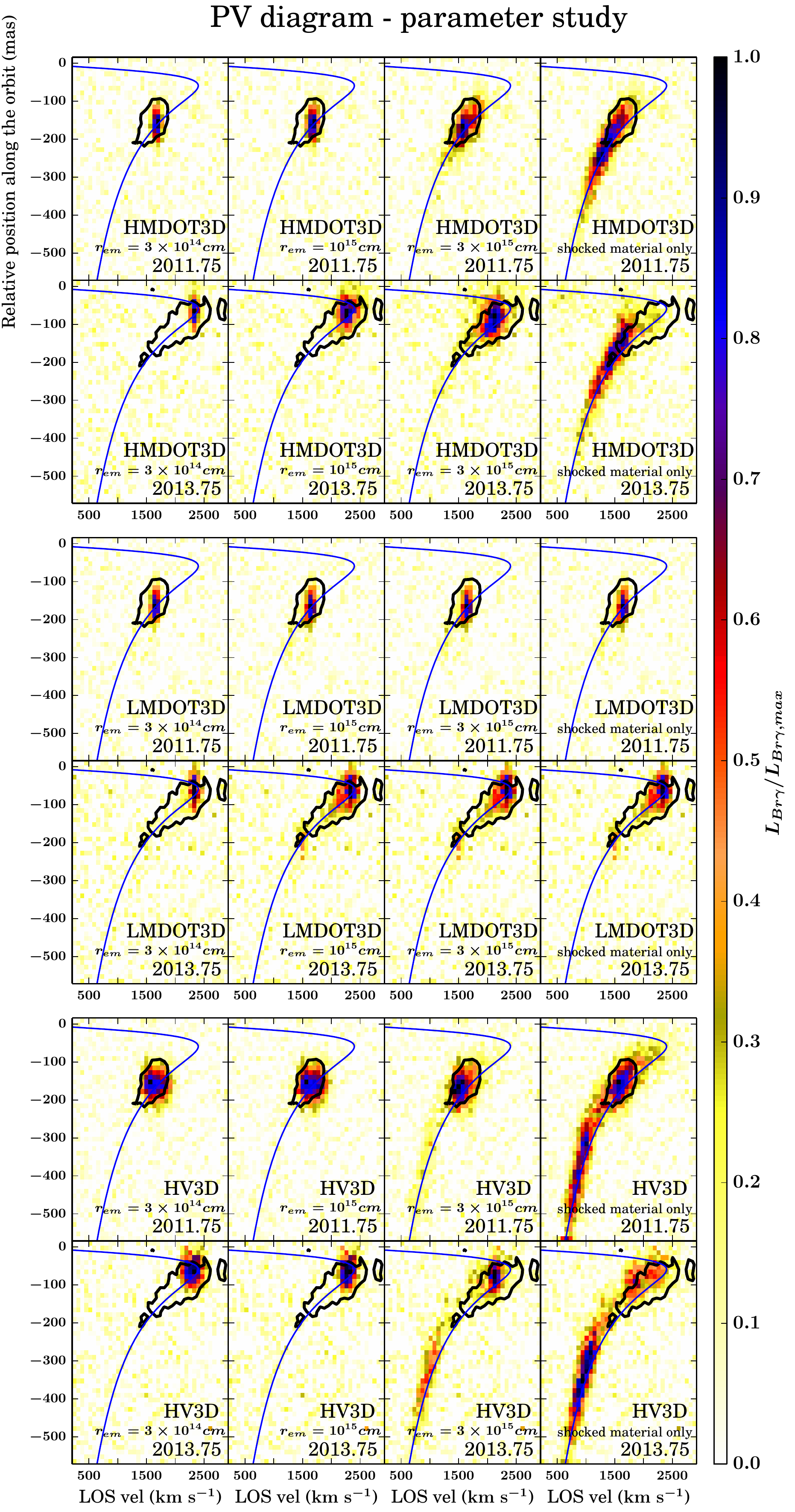}
\caption{Position-Velocity diagrams for our parameter study. The different panels show the simulated PV diagrams for different assumptions on the inner emitting radius $r_\mathrm{{em}}$. For every panel, the luminosity per bin is scaled to the maximum one. The black contours show the position and extent of the observed G2.
}\label{pvparam}
\end{center}
\end{figure}

Fig. \ref{pvparam} shows the PV diagrams for our parameter study. In the case of model HMDOT3D, G2 looks too elongated when only the shocked wind material is considered, while a reasonable match to observations could eventually be reached in the case of $r_\mathrm{{em}}>3\times10^{15} \;\mathrm{cm}$. Model LMDOT3D is instead producing a too compact emission for every assumption on $r_\mathrm{{em}}$. HV3D can instead result in a bimodal distribution in the PV diagrams, when looking at the emission of the shocked material only. For HV3D, the separation between the two simulated emission spots is not large enough to match the observed position of G2 and G2t on the orbit (see Fig. \ref{pvstandard}), but motivated our attempt to model both components with a single wind source \citep{Ballone16}.

\begin{figure}
\begin{center}
\includegraphics[scale=0.45]{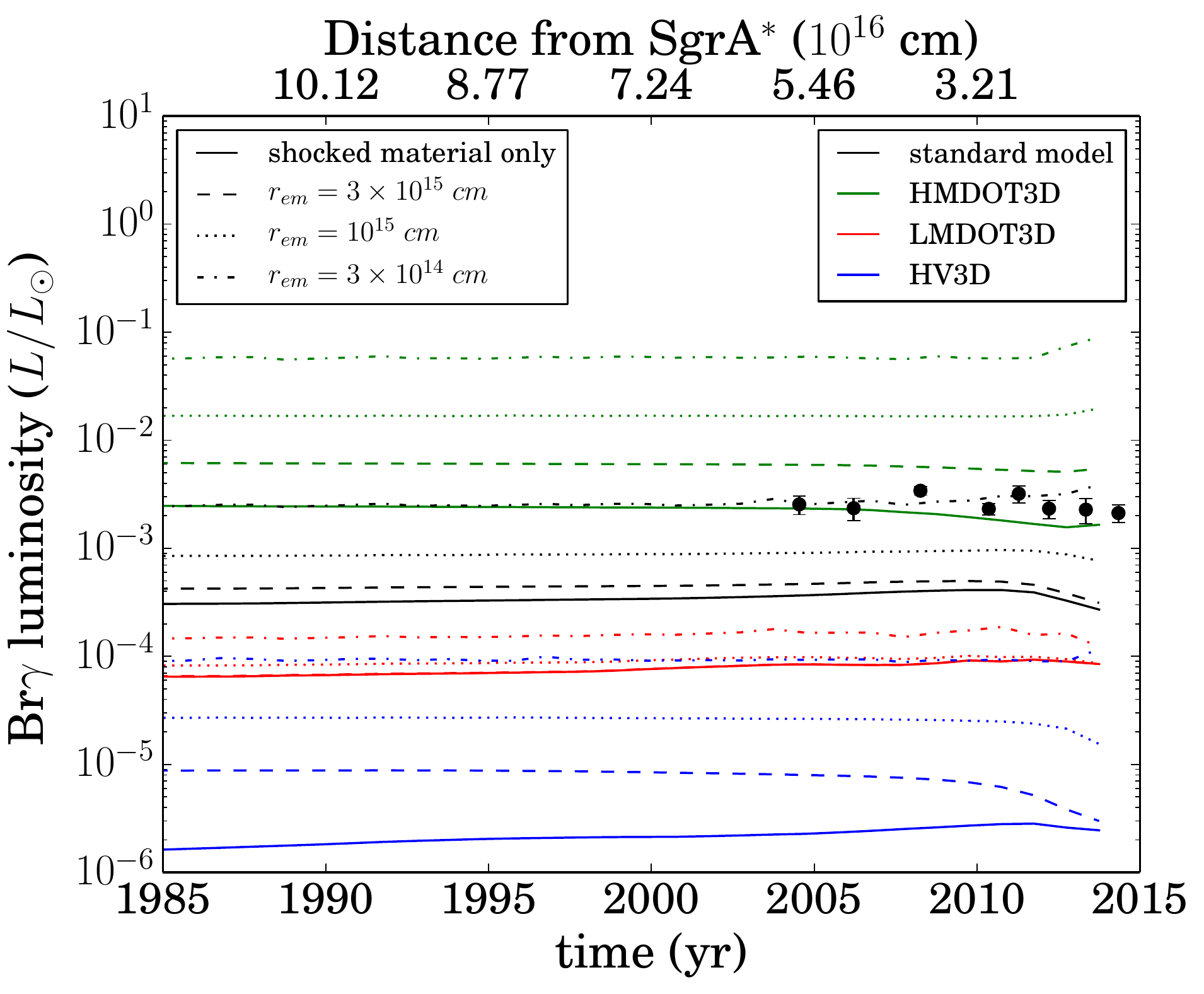}
\caption{Luminosity comparison for our simulations. The different colours show different models, the different linestyle refers to different assumptions for the inner emitting radius $r_\mathrm{{em}}$. The black points show the luminosities observed by \citet{Pfuhl15}.
}\label{lumall}
\end{center}
\end{figure}

When looking at the luminosity evolution in Fig. \ref{lumall}, while models LMDOT3D and HV3D have a too low luminosity \citep[roughly confirming the trends found in][]{Ballone13}, model HMDOT3D is matching the observations when the shocked-material only is considered, while it is a factor $\approx 2$ too luminous when $r_\mathrm{{em}}=3\times10^{15} \;\mathrm{cm}$ is adopted. The first evident effect is that lower mass-loss rates or higher velocities produce globally lower luminosities. This is simply explained by Eq. \ref{conteq} and \ref{lumform}, showing that the luminosity is proportional to the integral of $\rho_\mathrm{w}^2$. and that $\rho_\mathrm{w}$ is directly proportional to the mass-loss rate and inversely proportional to the outflow velocity. So, on a 0-th order, outflows with lower mass-loss rates and/or higher velocities are less dense (even in their shocked part) and have a lower emission measure, and vice versa. For any fixed model, a varying contribution is also given by the free-wind region, depending on the choice of $r_\mathrm{{em}}$. This result, however, is in contradiction with what has been found by \citet{Ballone13} with 2D simulations, where the shocked material was dominating the total luminosities close to pericentre. This is mainly explained by the poor resolution of the present simulations, as discussed in Sec. \ref{seclum}. As a consequence, we conclude that the absolute values of the calculated luminosities must be taken as lower limits, while the structure in the PV diagrams is a more solid and stable diagnostic tool.

\section{Discussion}\label{secdisc}

For a critical interpretation of the results presented in the previous section, a discussion of a few issues concerning the calculation of the Br$\gamma$ luminosity is needed.

The first issue is clearly visible in Fig. \ref{pvrem} and \ref{pvparam}: in order for this scenario to reproduce the size of G2 in the position-velocity space, most of the inner unperturbed part of the outflow must be neutral (hence, dark in recombination line emission; $r_\mathrm{{em}}\gtrsim 3 \times 10^{15} \; \mathrm{cm}$). Understanding whether this is actually the case would require a solid knowledge and treatment of the source of gas ionization. A full integration of radiation in the hydrodynamical simulation is needed, but the current simulations were already extremely time consuming and a further coupling with a radiative transfer scheme would make them unfeasible. On top of this, however, most of the available studies make use of simplified and (sometimes dramatically) different assumptions on the flux of ionizing photons reaching G2. In section \ref{secion} we present a very basic calculation of the amount of free-wind region that is actually ionized by Lyman-continuum (Lyc) photons from the surrounding stars. Such estimate contains several caveats, hence it does not have particularly strong physical basis; nonetheless, it clearly shows how the contribution of the free-flowing region can be severely dependent on the different assumptions on the Lyc photons flux on G2.

Another issue concerns the effect of numerics in the simulations on the resulting Br$\gamma$ luminosity and, particularly, on the luminosity curves in Fig. \ref{lumall}. We show in section \ref{seclum} that the absolute value of the total luminosity can be significantly affected by the resolution and by the geometry and symmetry used for the simulation. From this, we can conclude that the PV diagrams are more stable diagnostics, compared to luminosity curves, since they essentially represent the relative contribution to the total luminosity from different regions of G2.

After such needed discussions, the second part of this section tries to compare our work to what has been done by other authors (section \ref{seccomp}) and to give our model a more physical context, with a focus on the possible nature of the central source (section \ref{nature}) and of the advantages and disadvantages of this scenario, compared to the ``diffuse cloud'' one (section \ref{secprocon}).

\subsection{Ionization of the outflow}\label{secion}

In this section we try to estimate the contribution to the total luminosity of the free-flowing region of any wind in the Galactic Centre. The following calculation is based on the assumption that the ionization of the gas fully comes from UV photons from the nearby young stars \citep[see Section 1 and][]{Gillessen12}. Unfortunately, the flux of ionizing photons reaching G2 is not well known, so we decided to stick to a very simplified analytical calculation (see section \hyperref[sssec:num1]{4.1.1} for a discussion about its limitations). Its main purpose consists in justifying the need of $r_\mathrm{{em}}$ as a free parameter in the analysis of our hydrodynamic simulations.

The derivation is based on equating, in a one dimensional fashion, the rate of UV ionizing (Lyc) photons isotropically penetrating a spherical (``naked'' free-flowing) region, whose density scales as $1/r^2$, to the rate of recombinations occurring within the latter. In this way, we get the number of atoms in a free-wind shell needed to ``consume'' all the ionizing photons reaching G2. The thickness of this shell is depending on the total size of the free-wind region and, of course, on the amount of available Lyc photons.

The inferred equation is

\begin{equation}\label{ioninit}
\phi\left(\frac{R_{\mathrm{out}}}{D}\right)^2\approx\int_{R_{\mathrm{in}}}^{R_{\mathrm{out}}}\alpha_{\mathrm{rec}}n_\mathrm{e}n_\mathrm{i}4\pi r^2dr
\end{equation}

where $\phi$ is the rate of emitted ionizing photons and $D$ is the distance of G2 from the ionizing source. However, the value of these two latter quantities is not very well constrained and one must assume there is more than one emitting source. In the following, $\phi/4\pi D^2$ will simply be the flux of ionizing photons on G2 and we will consider different numbers used in previous calculations by different authors.
$R_{\mathrm{out}}$ and $R_{\mathrm{in}}$ are respectively the outer and inner radius of the ionized shell. $\alpha_{\mathrm{rec}}$ is the total recombination coefficient and we assumed $\alpha_{\mathrm{rec}}=2.59\times10^{-13}\; \mathrm{cm^3 s^{-1}} $, i.e. the value for case B recombination for pure hydrogen at $T=10^4\; \mathrm{K}$ \citep[][page 22]{Osterbrock06}. $n_\mathrm{e}$ and $n_\mathrm{i}$ are the number densities of the electrons and ions - respectively - in the gas and $r$ is the distance from the source. For a $1/r^2$ density profile,

\begin{equation}\label{denprof}
n_\mathrm{e}n_\mathrm{i}\approx \frac{\rho^2}{\mu_\mathrm{e}\mu_\mathrm{i}m_\mathrm{H}^2}\approx\frac{\dot{M}_\mathrm{w}^2}{16\pi^2v_\mathrm{w}^2\mu_\mathrm{e}\mu_\mathrm{i} m_\mathrm{H}^2r^4},
\end{equation}

where $\dot{M}_\mathrm{w}$ and $v_\mathrm{w}$ are the mass-loss rate and velocity of the wind, respectively, $\mu_\mathrm{e}=1.17$ and $\mu_\mathrm{i}=1.29$ are the electron and ion mean weight (for solar metallicity) and $m_\mathrm{H}$ is the hydrogen mass.

So, substituting $n_\mathrm{e}n_\mathrm{i}$ in Eq. \ref{ioninit} and solving the integral, we can get the inner radius $R_{\mathrm{in}}$ for which there is a balance between the rate of incoming ionizing photons and the rate of recombinations, over the whole volume:

\begin{equation}\label{inner1}
R_{\mathrm{in}}=\left[\frac{\phi}{D^2}\frac{4\pi v_\mathrm{w}^2\mu_\mathrm{e}\mu_\mathrm{i} m_\mathrm{H}^2 R_{\mathrm{out}}^2}{\alpha_{\mathrm{rec}} \dot{M}_\mathrm{w}^2}+\frac{1}{R_{\mathrm{out}}}\right]^{-1}.
\end{equation}

The total volume of the free-wind region changes as the source moves along the orbit and encounters a higher and higher external pressure. In this case, we assume that the outer radius $R_{\mathrm{out}}$ is just the stagnation radius given by the atmosphere's thermal pressure only (we hence neglect any anisotropic pressure contributions)

\begin{equation}\label{stag}
R_{\mathrm{out}}=\left[\frac{\dot{M}_\mathrm{w}v_\mathrm{w}}{4\pi P_{\mathrm{amb}}}\right]^{1/2}.
\end{equation}

Substituting Eq. \ref{stag} in Eq. \ref{inner1}, we get

\begin{equation}
R_{\mathrm{in}}=\left[\frac{\phi}{D^2}\frac{v_\mathrm{w}^3\mu_\mathrm{e}\mu_\mathrm{i} m_\mathrm{H}^2}{\alpha_{\mathrm{rec}} \dot{M}\mathrm{_w} P_{\mathrm{amb}}}+\left(\frac{4\pi P_{\mathrm{amb}}}{\dot{M}_\mathrm{w}v_\mathrm{w}}\right)^{1/2}\right]^{-1}.
\end{equation}

For our choice of the atmosphere (see Eq. \ref{denatm} and \ref{tematm}) the ambient thermal pressure is varying with radius and so will the inner and outer radii do:

\begin{equation}\label{inner2}
\begin{split}
R_{\mathrm{in}}& \approx3\times10^{15}\left[1.691\times10^{-2}\frac{\phi_{50}}{D_{\mathrm{pc}}^2}\frac{v_{\mathrm{w,50}}^3d_{\mathrm{BH,peri}}^2}{\dot{M}_{\mathrm{w,-7}}}+\right.\\
&\left. +44.099\left(\frac{1}{\dot{M}_{\mathrm{w,-7}}v_\mathrm{{w,50}}}\right)^{1/2}\frac{1}{d_{\mathrm{BH,peri}}}\right]^{-1} \mathrm{cm},
\end{split}
\end{equation}

where we expressed the rate of ionizing photons in units of $10^{50} \mathrm{s^{-1}}$, D in units of pc, the wind's mass-loss rate $\dot{M}_\mathrm{w}$ in units of $10^{-7}\;M_{\odot} \;\mathrm{yr^{-1}}$, its velocity $v_\mathrm{w}$ in units of $50 \;\mathrm{km\; s^{-1}}$ and the distance from SgrA* $d_{\mathrm{BH,peri}}$ in units of the pericentre distance.

We performed the calculation for $v_\mathrm{w}=50 \;\mathrm{km\;s^{-1}}$ and $\dot{M}_\mathrm{w}=10^{-7},10^{-6}\; M_{\odot} \mathrm{yr^{-1}}$, assuming five different fluxes of UV photons: 

\begin{itemize}
\item In the first case, which we will call \textbf{SB}, we assumed the numbers used by \citet{Scoville13}, i.e. $\phi_{50}=1$ and $D_{\mathrm{pc}}=1$. This assumption is equivalent to having a single O5 star at a constant distance of 1 pc. 
\item In the second case, \textbf{MLlow}, we took numbers from \citet{Murray-Clay12} for the entire central parsec; these numbers are (more or less) matching the values provided in \citet{Martins07} and \citet{Genzel10}. In particular, they assume $\phi_{50}=10^{0.8} \simeq 6.31$ for $D_{\mathrm{pc}} =1$. 
This is a lower estimate for the UV flux given by these authors. 
\item \citet{Murray-Clay12} also took into account the concentration of the S-stars (of spectral class B) within the very central region of the Galactic Centre. They estimate these stars to produce a total $\phi_{50}=0.2$, but for a region of $D_{\mathrm{pc}}\simeq 6\times10^{-3}$. This is their higher estimate 
and we refer to it as \textbf{MLhigh}.

\item In the fourth case, \textbf{Sh04}, we assumed the flux used by \citet{Shcherbakov14} for the position of the cloud in the year 2004. We consider the values derived by this author as the most reasonable ones, since they are obtained calculating the contribution of the main Wolf-Rayet stars in the young cluster, exactly taking into account their positions, from \citet{Paumard06} and \citet{Lu09}, and their temperatures and luminosities, from \citet{Martins07}. In 2004, $F_{\mathrm{UV}}=3\times10^4 \mathrm{\;erg\;s^{-1}\;cm^{-2}}$. If we crudely divide this value by the ionization energy of the hydrogen atom, we get the number flux of ionizing photons $\phi_{50}/D_{\mathrm{pc}}^2\simeq131$.
\item In the last case, \textbf{Sh14}, we assumed the flux assumed by \citet{Shcherbakov14} at G2's pericentre, namely $F_{\mathrm{UV}}=5.7\times10^4 \mathrm{\;erg\;s^{-1}\;cm^{-2}}$. Close to pericentre, the flux increases due to the contribution of the star S2. Dividing by the ionization energy of the hydrogen atom, we get $\phi_{50}/D_{\mathrm{pc}}^2\simeq249$.

\end{itemize}

\begin{figure}
\begin{center}
\includegraphics[scale=0.52]{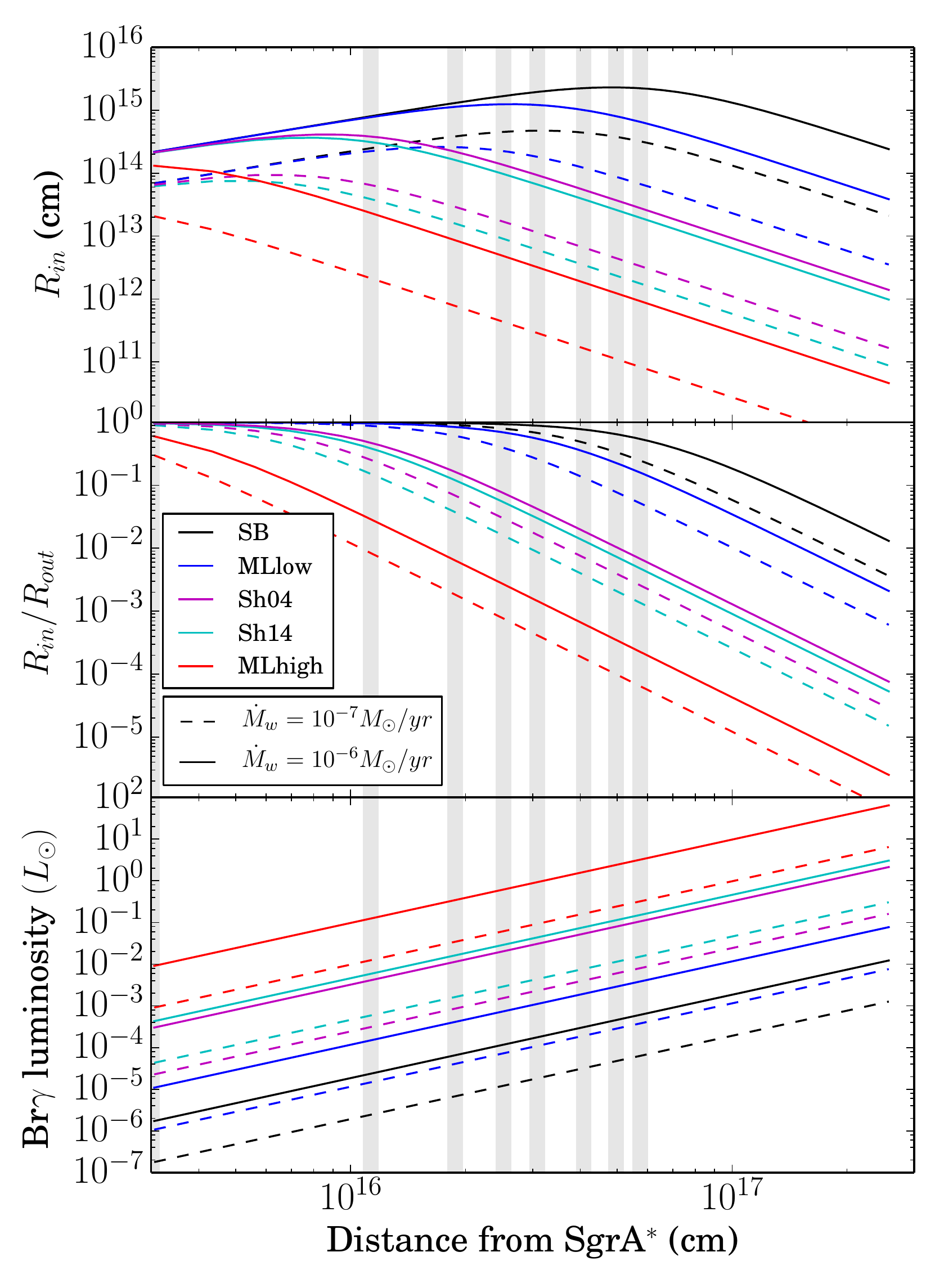}
\caption{Results of our analytical calculation for the ionization of the free-wind region discussed in Sec. \ref{secion}. The upper panel shows the absolute value of the inner radius $R_{\mathrm{in}}$, while the central panel shows the ratio between the inner ($R_{\mathrm{in}}$) and outer ($R_{\mathrm{out}}$) radii of the ionized shell. The luminosity of the spherical free-flowing ionized shell is plotted in the lower panel. The dashed and solid lines show the results for $\dot{M}_\mathrm{w}=10^{-7},10^{-6}\; M_{\odot} \mathrm{\;yr^{-1}}$, respectively (the wind velocity is $v_\mathrm{w}=50 \mathrm{km\;s^{-1}}$ for both calculations). Different colours show the results for different assumptions on the flux of ionizing photons in the Galactic Centre: $(\phi/D^2)_\mathrm{SB}=1$, $(\phi/D^2)_\mathrm{MLlow}=6.31$, $(\phi/D^2)_\mathrm{Sh04}=131$, $(\phi/D^2)_\mathrm{Sh14}=249$, $(\phi/D^2)_\mathrm{MLhigh}=5560$. The grey vertical bands correspond to $d_{\mathrm{BH}}$ of G2 for the different times of its monitoring, i.e. years 2004.25, 2006.25, 2008.25, 2010.25, 2011.25, 2012.25, 2013.25.
}\label{ioniz}
\end{center}
\end{figure}

In Fig. \ref{ioniz} we plot the results of our simple analytical calculation. As visible in the central panel, the size of $R_\mathrm{{in}}$ first increases with G2 getting closer to SgrA* and then decreases at smaller distances. This is the result of two competing effects, i.e. the decrease of available ionizing photons and the increasing density (and number of recombinations) in the outer layer of the free-wind region with the shrinking of the stagnation radius. These two different branches are mathematically visible in Eq. \ref{inner1} and \ref{inner2}, as asymptotic branches $\propto R_{\mathrm{out}}^{-2}\propto d_{\mathrm{BH}}^{-2}$ for large distances and $\propto R_{\mathrm{out}}\propto d_{\mathrm{BH}}$ for small ones (see also the upper panel in Fig. \ref{ioniz}). It is also interesting to note that the transition between these two branches moves to larger values of $d_{\mathrm{BH}}$ for smaller values of $\phi/D^2$. On the other hand, the lower panel of Fig. \ref{ioniz} shows that the Br$\gamma$ luminosity is a monotonic function of $R_{\mathrm{out}}$ and $d_{\mathrm{BH}}$. This is easily understandable from Eq. \ref{ioninit}: as the Br$\gamma$ luminosity is directly proportional to the number of recombinations (i.e., the right hand side of the equation), it is also $\propto R_{\mathrm{out}}^2$.

As just described, the evolution of $R_{\mathrm{in}}$, $R_{\mathrm{in}}/R_{\mathrm{out}}$ and the Br$\gamma$ luminosity, as a function of the distance from the black hole is a direct result of the previous equations, hence of our assumptions. The important result is that these quantities strongly depend on $\phi/D^2$, spanning orders of magnitude for $d_{\mathrm{BH}}$ corresponding to the observations (grey vertical bands). To this end, instead of the derived $R_\mathrm{{in}}$, we adopted $r_\mathrm{{em}}$ as a free parameter in the hydrodynamic simulations in Section \ref{secresults}.

\subsubsection{Caveats}\label{sssec:num1}

The presented calculation contains a large number of approximations.

First of all, we neglect the role of shielding due to the dense shocked material around the free-wind region; this can lead to substantially lower ionization in the free-wind region.

Another extreme simplification is related to the assumed spherical symmetry of the calculation. Our Eq. \ref{ioninit} is based on the idea that G2 is hit at $R_{\mathrm{out}}$ by $\phi/D^2$ photons coming from all directions, which is, of course, not the case. In reality, any surrounding young star will contribute to the illumination of G2 in a different way, dependent on its spectral class and position, hence making the flux not isotropically distributed on G2's surface. In addition to this, the pressure contributions (particularly the ram and tidal ones) shaping the free-wind region will make the free-wind surface asymmetric \citep[for a discussion of the physics of stellar winds in the Galactic Centre, see Sec. \ref{secresults} and][]{Ballone13,Ballone16,Christie16}.

Another caveat is related to the assumptions that ionizations and recombinations both occur istantaneously and that all the photons impinging on $R_{\mathrm{out}}$ are totally absorbed by the free-wind region. However, as shown in \citet{Mapelli15}, the timescales for these two processes might be very different. In our case, the recombination timescale is

\begin{equation}
t_\mathrm{rec}(r)=\frac{1}{\alpha_\mathrm{{rec}}n_\mathrm{i}(r)}\approx5\times10^5\frac{r^2_{14}v_\mathrm{w,50}}{\dot{M}_\mathrm{w,-7}} \;\mathrm{s},
\end{equation}

where $r_{14}$ is $r$ in units of $10^{14} \;\mathrm{cm}$. The ionization timescale is

\begin{equation}
t_\mathrm{ion}=\frac{4\pi D^2}{\sigma_\mathrm{H}\phi}\approx 2\times10^5\frac{D^2_\mathrm{pc}}{\phi_{50}} \;\mathrm{s},
\end{equation}

where $\sigma_\mathrm{H}\simeq6.3\times10^{-18}\;\mathrm{cm}^2$ is the cross section for neutral hydrogen and photons with energy $13.6 \;\mathrm{eV}$. An equilibrium between ionizations and recombinations can be assumed if $t_\mathrm{rec}=t_\mathrm{ion}$, which does not always hold for our assumptions. We also ignore that a certain number of photons (i.e., those passing through the outer envelope tangentially) might escape the free-wind region before ionizing any atom.

Finally, other physical processes could be important as well, such as collisional ionization from the wind \citep[as already shown by][]{Scoville13} or absorption of Lyc photons by the dust embedded in G2.

As already stated, the number of caveats listed here does not allow a strict use of the calculation for the modeling of G2's emission. Nonetheless, it powerfully shows that the contribution of the inner part of the outflow to the Br$\gamma$ luminosity of G2 is not trivial.\\

\begin{figure}
\begin{center}
\hspace*{-0.65cm}
\includegraphics[scale=0.46]{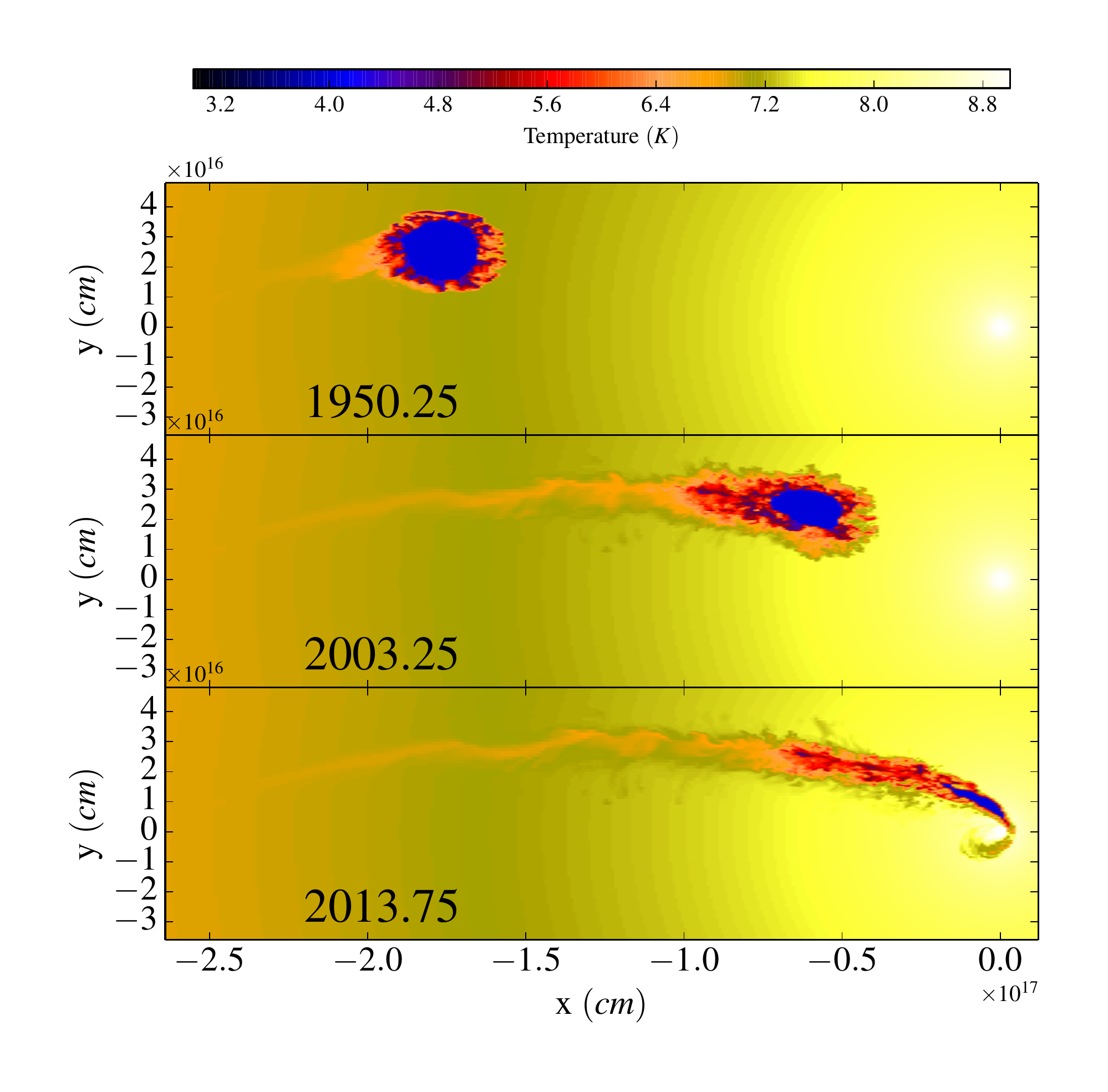}
\caption{Temperature maps for the standard model. The domain plotted is a slice at z=0.
}\label{tempmap}
\end{center}
\end{figure}

\begin{figure*}
\begin{center}
\includegraphics[scale=0.48]{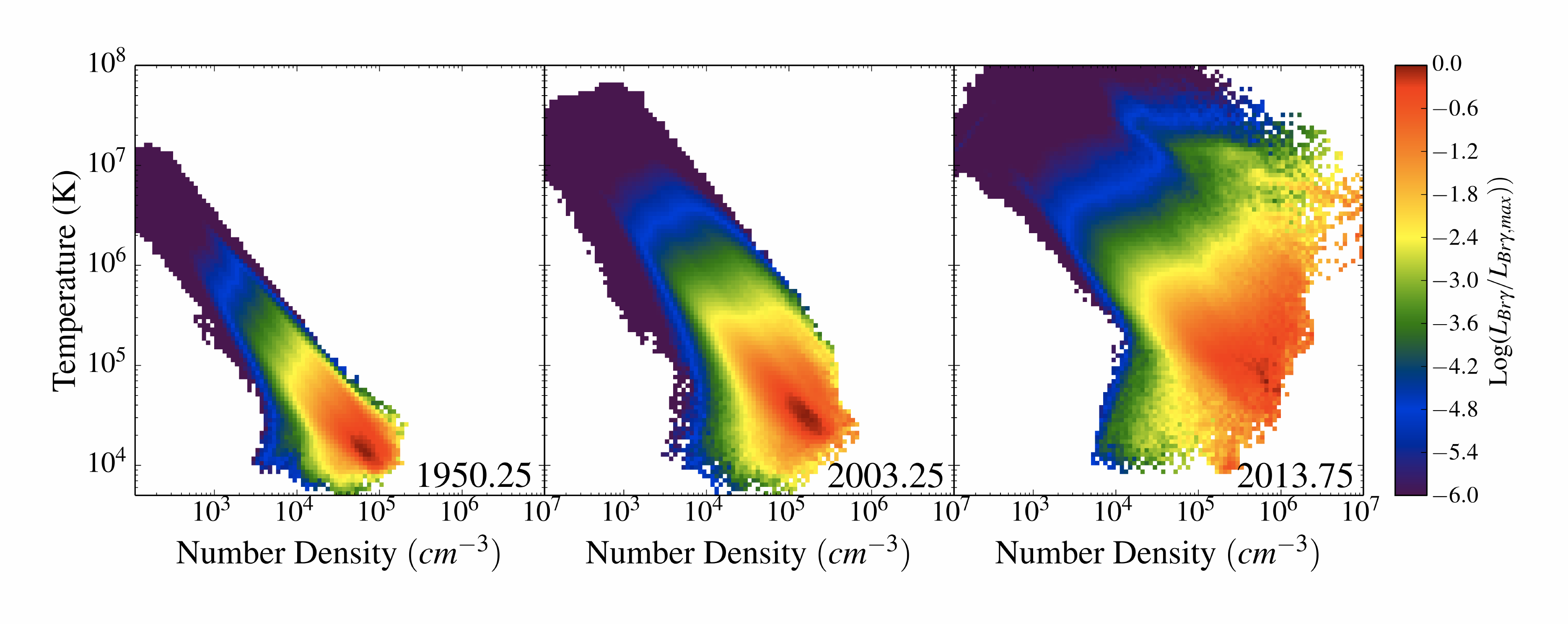}
\caption{Phase plots for our standard model. Only the shocked material is included.
}\label{phasestandard}
\end{center}
\end{figure*}

\begin{figure}
\begin{center}
\includegraphics[scale=0.42]{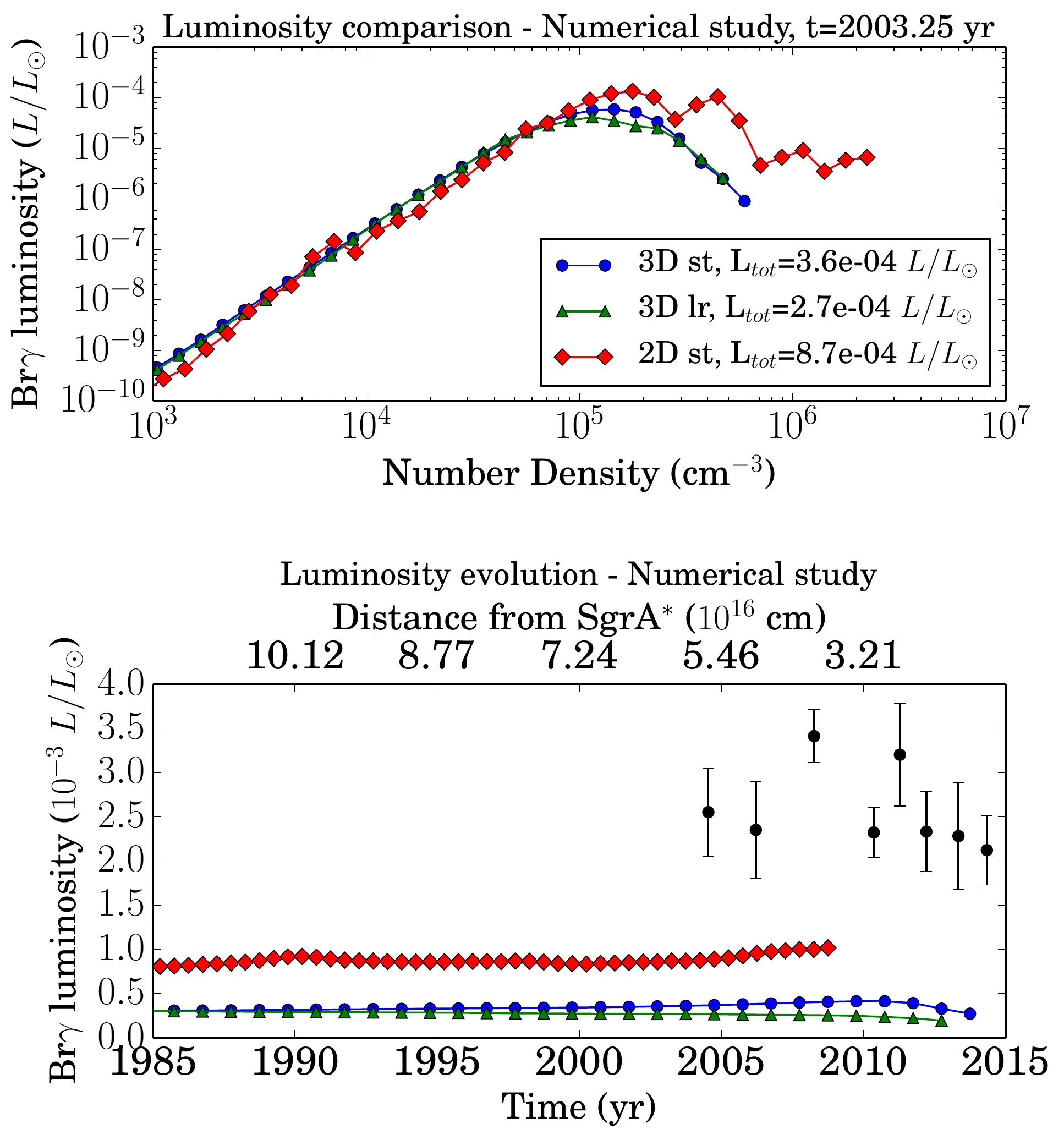}
\caption{Luminosity comparison for our numerical study. Upper panel: contribution to the total luminosity by different density bins, in year 2003.25, for the standard model (blue circles), stLOWRES (green triangles) and st2D (red diamonds). Lower panel: luminosity evolution, close to pericentre, for the standard model, stLOWRES and st2D. Colours and symbols are the same as in the upper panel. The black points with error bars show the luminosities observed by \citet{Pfuhl15}.
}\label{resstudy}
\end{center}
\end{figure}

\subsection{Resolution and numerical issues}\label{seclum}

The constraining power of the absolute value of the Br$\gamma$ luminosity has to be reconsidered, after the systematic study of the present 3D simulations. 

First of all, the shocked material has a very filamentary nature; hence, if the filaments are not properly resolved, the density of the shocked material is reduced significantly. 

Furthermore, as shown in Fig. \ref{tempmap} and \ref{phasestandard}, the shocked material is efficiently mixing with the atmosphere, moving to higher temperatures. In particular, as visible in Fig. \ref{tempmap}, at early stages (e.g., in year 1950.25) the material closest to the free-wind region is at temperatures of around $10^4$ K, i.e. the temperature of the injected material. However, the mixing becomes faster and faster as the source reaches its pericentre: at 2013.75, most of the shocked material is immediately increasing its temperature and a relatively small fraction is at temperatures below $10^5$ K. The evolution of the phase plots for the shocked material in Fig. \ref{phasestandard} might be misleading, since mixing with lighter material should also reduce its density with time. However, the diagram shows that the most luminous material increases its density (i.e., it moves to the right of the plot) as the source approaches pericentre; this is simply due to the fact that the outflow moves faster and it encounters higher density/pressure material on its way to the black hole. So, it is compressed more and reaching higher densities, as it gets closer to SgrA* (see also Fig. \ref{denmap}). On the other hand, the inner atmosphere is also hotter, hence the mixing leads to a large spread of the shocked material in the density-temperature phase space. 

The mixing in our simulations is resolution dependent, since its nature is partly numerical. This has already been shown in \citet{Schartmann15} for the diffuse cloud scenario. In the case of the present outflow model, mixing and resolution limitations are significant all along the orbital evolution, given the highly filamentary nature of the emitting material, and the effect of poor resolution is even less predictable. 

The upper panel of Fig. \ref{resstudy} shows histograms for the luminosity of our standard model as a function of the density of the emitting material, for 2003.25 (i.e., the central panel of Fig. \ref{phasestandard} collapsed along the temperature axis). The same histogram is plotted also for the simulation stLOWRES (the same as our standard model, but with half of the resolution) and for the simulation st2D (the same as the standard model, but in a 2D cylindrical fixed grid; see Table \ref{parsimul}). The luminosity distribution peaks\footnote{The luminosity is given by the product between the density and the volume occupied by gas at that density. The latter is a decreasing function of the density, explaining the presence of a peak in the histogram.} around densities of roughly $10^{-19} \mathrm{\; g \;cm^{-3}}$ for the outflow parameters of these three simulations. Though small, some discrepancy occurs between the two 3D simulations at different resolutions, particularly close to the peak of the distribution. This can account for the difference in the luminosity evolution, close to pericentre (see the lower panel of Fig. \ref{resstudy}), between our standard model and model stLOWRES. The effect of resolution on the luminosity evolution is similar to the one visible in the resolution study of \citet{Schartmann15}.

A way more significant difference occurs, instead, between the luminosity curves of our standard model and its two-dimensional counterpart st2D. As already discussed in \citet{Ballone13}, simulations in 2D cylindrical coordinates suffer from some intrinsic numerical issues: in particular, the accumulation of material towards $R=0$, due to the tidal field of the SMBH, is artificially enhanced by the cylindrical symmetry and by the necessary reflective boundary conditions close to the axis of symmetry. For this reason, in order to compare with our standard model, we removed all the material at $R<1.25\times10^{14} \; \mathrm{cm}$. However, still a significant contribution to the luminosity of st2D comes from densities higher than $2\times 10^{-19} \mathrm{\; g \;cm^{-3}}$, while this is not the case for the 3D standard model, showing that the artificial compression towards the axis of symmetry might have an effect on even larger distances from the axis. Furthermore, given the cylindrical symmetry, in the calculation of the luminosity the volume of every cell is obtained by a rotation of the cell around the $z$ axis (i.e., every cell has to be thought as a 3D annulus). As a consequence, the dense Rayleigh-Taylor fingers forming immediately around the free-wind region might have a larger volume filling factor, compared to their 3D more realistic counterparts. All in all, there is a factor $\approx 2.5$ difference between the standard model and st2D, which forces us to also reconsider the luminosity curves obtained in the preliminary study of \citet{Ballone13}.

\subsection{Comparison with previous works}\label{seccomp}

In addition to the adopted dimensionality and coordinate system of the simulations, there are few additional differences between the simulations in \citet{Ballone13} and the present ones. The first one is that the orbit has been updated from the one derived by \citet{Gillessen13a} to the most-recent one derived by \citet{Gillessen13b}. The most up-to-date orbit has an orbital time and an apocentre distance that are roughly a factor of two larger than the one of the previous 2D simulations. This had the unfortunate effect of increasing the computational domain and double the integration time of our simulations, making the new simulations even more computationally expensive than previously expected.

Further, compared to \citet{Ballone13}, the mass-loss rate of our standard model has increased by roughly a factor 5. This has been induced by the need of matching the PV diagrams shown in Fig. \ref{pvstandard}. In fact, the parameters of the best model in \citet{Ballone13} are roughly corresponding to the LMDOT3D model described in Sec. \ref{secparam}, which is not able to match the size of G2 in the observed PV diagrams (see Fig. \ref{pvparam}). The increase in the mass-loss rate of the best model is probably due to the more accurate comparison performed here, as well as to intrinsic differences between 2D and 3D simulations. Additionally, there are major differences in the absolute value of the luminosity, as discussed in Sec. \ref{seclum}, due to differences in the resolution and perhaps in intrinsic differences between 3D cartesian and 2D cylindrical coordinates.

The choice of starting the simulations at apocentre makes the present results also very different from the ones in \citet{DeColle14}. In fact, the $\approx200 \; \mathrm{yr}$ evolution of our models \citep[compared to the 3 and 20 $\mathrm{yr}$ chosen by][]{DeColle14} leads to a much more extended distribution of gas, as a result of the prolongated stripping of the RTI filaments of shocked wind. This larger filling volume is fundamental for matching the observed PV diagrams. However, no major instability forms in the simulations of \citet{DeColle14}, probably as a result of the too short evolution time of their models.

Major differences between our simulations and the ones in \citet{DeColle14} also arise around pericentre, where the bow-shocks in their simulations - particularly those starting 3 years before pericentre - are becoming broader and underdense after the pericentre passage. This might be a consequence of their more sophisticated treatment of radiative cooling. The difference might also arise from the fact that, for those simulations, \citet{DeColle14} did not artificially stabilize their atmosphere. This is allowing to compute the bow shock dynamics more properly, but it has the side effect of allowing the atmosphere to become convectively unstable \citep[as clearly visible in Fig. 1 of][]{DeColle14}.

Our work is also complementary to that by \citet{Zajacek16}. In this work, the evolution of the stellar wind shock is studied by means of the analytic solution of \citet{Wilkin96}. Such estimates have the advantage of having a simple but ``linear'' description of the interaction between the wind and the surrounding atmosphere; however, they lack more complex hydrodynamic processes that already arise from our simulations, even with our relatively simple physical treatment.

We must also point out that, besides lacking the detailed procedure to mock the instrumental effect on the processing of the simulation, the mock Br$\gamma$ maps and the PV diagrams shown in \citet{Ballone13}, \citet{Gillessen13b} and \citet{DeColle14} include all the outflow material present in the simulations. This choice is arbitrary, since it depends on how much of the free-wind region is actually resolved in the simulation, and can produce PV diagrams with Br$\gamma$ fluxes that are spanning several order of magnitudes, in evident inconsistency with the observations (compare to the upper panel of Fig. \ref{pvstandard}). Furthermore, as discussed in Sec. \ref{secion} and further on, the Br$\gamma$ luminosity of a $1/r^2$ density distribution depends on how much of it is actually ionized. Our more detailed post-processing of the simulation clearly shows that a more careful interpretation of the results must be applied, when dealing with this scenario.

Finally, this (and the previously mentioned) works focused on reproducing only G2, while the study presented in \citet{Ballone16} tries to use the same model to simultaneously explain the presence of G2 and the following G2t. In this regard, even considering the weak constraining power of the Br$\gamma$ luminosity, the present study shows that there should be a significant effect of the outflow parameters on the total luminosity of the shocked gas. Hence, the present standard model and the one in \citet{Ballone16} are mutually exclusive. The model described in \citet{Ballone16} has the advantage of being able to give G2 and G2t a common origin (even though the physical link between these two components is yet to be fully proven), but it has to be regarded as a proof of concept study and fine tuning of the model parameters is necessary to meet all observational constraints. The present standard model, on the other hand, is only able to reproduce G2, but it seems to have a Br$\gamma$ luminosity that is closer to the observed one.

\subsection{On the nature of the source}\label{nature}

As already shown in \citet{Gillessen12}, the spectral properties of G2 exclude its association with a massive star, such as the S-stars. At the same time, the mass loss rates of our models are all too high for typical winds of low-mass stars in their main sequence phase. 

Low-mass stars in their asymptotic giant branch or red giant phases might have comparable high mass loss rates \citep[see, e.g.,][]{Whitelock16}. Stars in these phases have a giant envelope, that usually leads to outflow velocities of the order of their escape velocities, i.e., few tens of km/s. This is indeed the case for the standard model. However, those stars would appear too bright in $K_s$ band, compared to G2, and this possibility can be excluded.

The most appealing possibility is that the source is instead a young star, such as a T Tauri \citep[see also][]{Scoville13,Ballone13}. These objects are also producing winds, but they have much lower luminosities in $K_s$ band \citep[see discussion in][]{Scoville13}. However, the parameters of the present 3D standard model ($\dot{M}\mathrm{_w=5\times 10^{-7} M_{\odot} \;yr^{-1}}$ and $v\mathrm{_w = 50 \;km\, s^{-1}}$) are somehow at the extreme end of the observed ranges for T Tauri's winds, which are $\dot{M}\mathrm{_w=[10^{-12},10^{-7}] \;M_{\odot} \;yr^{-1}}$ and $v\mathrm{_w = [50,300] \;km/s}$ from the observations \citep{White04}. Given the short evolution time of our models ($\approx 200\;\mathrm{yr}$), the standard model parameters could still correspond to a phase of exceptionally higher mass-loss. Indeed, there is a well established correlation between mass accretion and outflow rates for T Tauri objects, possibly being the consequence of outflows launched from the proto-stellar accretion disk \citep[e.g.,][]{White04, Edwards06}. In such a crowded environment and given the high tidal field of the black hole, the accretion (and outflow) rates might be enhanced compared to the typical star forming regions. Extremely massive outflows have been discovered, as e.g. for the case of DG Tau \citep{Gunther09, White14}.

This problem can also be partially ``cured'' by assuming that the outflow is biconical, i.e., it is not occupying the full solid angle. As widely shown in literature, this is indeed a much more realistic assumption for the outflows from this kind of young stellar objects \citep[e.g.,][]{Torbett84}. In this case, Eq. \ref{stag} becomes

\begin{equation}\label{stagtheta}
R_\mathrm{{out, conical}}=\left[\frac{\dot{M}_\mathrm{w}v_\mathrm{w}}{4\pi (1-cos\theta_\mathrm{{open}}) P_\mathrm{{amb}}}\right]^{1/2},
\end{equation}

where $\theta_\mathrm{{open}}$ is the half opening angle of the outflow. So, for the same value of $R_\mathrm{{out}}$, in the case of a biconical outflow, $\dot{M}_\mathrm{w}$ can be a factor $(1-cos\theta_\mathrm{{open}})$ (i.e., up to a factor $\approx10^{-2}$ for half opening angles as small as $\approx 10^{\circ}$) smaller compared to the isotropic case tested here. As shown in Sec. \ref{secparam}, the stagnation radius is on a 0th order responsible for the size of the outflow; hence, to get sizes similar to the observed ones, lower mass-loss rates could be needed, compared to the ones found in our current simulations. However, the orientation of the biconical outflow with respect to the orbit is also likely effecting the distribution of the emitting material. This would add a further parameter to the present scenario and additional dedicated simulations would be needed to clarify this issue. 

\subsection{Advantages and disadvantages of a compact source scenario}\label{secprocon}

As pointed out by the present and previous studies \citep{Ballone13, DeColle14,Zajacek16,Ballone16}, the compact source scenario is a highly parametric model, which makes the results strongly dependent on the assumptions made. Its intrinsic properties also make its study numerically challenging. Occam's razor would then suggest us that a diffuse cloud scenario \citep[possibly without any hydrodynamical interaction with the accretion flow, as the one originally proposed by][]{Gillessen12} is to be preferred. However, more parameters can always offer more possibilities to reconcile the model and the observations. 

For example, \citet{Pfuhl15} showed that the total mass of the dust embedded in G2 is probably too low to make this component dynamically important. On the other hand, \citet{Witzel14} showed that the dust stays compact even close to pericentre, compared to its gaseous counterpart. It is not clear why this should happen in a diffuse cloud scenario. An outflow nature for G2 has the advantage of explaining both the extended (in position and velocity) nature of the gas component and the compactness of the dusty emission, if the latter is associated to a central young stellar object. 

Another open question is related to the high eccentricity of G2's orbit. This could be well explained by a formation of G2 in colliding winds in the disk, if G2 is a clump of diffuse gas \citep{Burkert12, Schartmann15,Calderon16}. Compared to stars, gas can more easily lose angular momentum (and energy) and the collision of stellar winds represents a very effective process, in this sense. However, the inner parsec is also very crowded with young stars (as young as T Tauri, see Sec. \ref{nature}) and the S-stars can have similar orbital semi-major axes and can reach similarly high eccentricities.

A connection to a star could then be possible. The binary merger model of \citet{Witzel14} could explain the dust properties and the high eccentricity of G2 \citep[as later shown by][]{Prodan15,Stephan16}, but so far completely neglected the existence of a significant gaseous component associated with it. Outflow models are often invoked to explain the latter (see Sec. \ref{intro}), but often they rely on - sometimes, too simple - analytical estimates. Despite the many limitations discussed in this section, our study represents the most complete attempt to include the several (often non-linear) processes involved in a compact source scenario and to compare to the observed properties of G2, e.g., by means of accurate mock PV diagrams.

Concerning the connection of G2 to G2t and G1 (see Sec. \ref{intro}), \citet{Guillochon14} showed that these objects might result from the stripping of the outer envelope of a giant star by the tidal field of SgrA*. Hydrodynamical simulations of tidal disruptions of stars by SMBHs indeed show that these events might lead to the formation of a bound debris, streaming towards the SMBH on highly eccentric orbits \citep[see also][]{Guillochon14b}. The fragmentation of such a streamer might have led to G1, G2 and G2t. The formation of multiple clumps in colliding winds \citep{Burkert12, Schartmann15,Calderon16} is also a very reasonable explanation. G1 and the G2+G2t complex have very similar orbital and emission properties. This naturally suggests a common or similar origin. Proving that they were all born at the same location is less straight-forward; for example, G2's pre-pericentre and G1's post-pericentre orbits do not coincide perfectly and have a very different apocentre position. To reconcile the two, some loss of energy and angular momentum could have occured - mainly at pericentre -, due to the interaction of these clumps and the surrounding atmosphere \citep{Pfuhl15, McCourt15, McCourt16, Madigan17}. However, \citet{Plewa17} showed that G2 is keeping its original orbit even after pericentre, excluding the latter hypothesis of a strong drag of the atmosphere on these clumps. G2 and G1 could still be related, but the new findings show that these objects did not have exactly the same orbit, before pericentre. As shown by the HV3D model presented here and by the one in \citet{Ballone16}, an outflow with low enough density can efficiently form a tail of stripped gas (with properties similar to the observed G2t), although the source keeps on moving (and losing new material) on a purely Keplerian orbit.
 
The state-of-the-art models on G2's nature are all able to reproduce some of G2's properties, but also show limitations or are unable to explain other observables. 
Additionally, the pericentre evolution of G2 in simulations for the diffuse cloud scenario \citep{Schartmann12,Anninos12,Schartmann15} and in our simulations look very similar and the comparison to mock PV diagrams \citep{Schartmann15} shows that both models might be reconciled with observations. Hence, no final conclusion can be drawn, yet. The smoking gun for understanding whether a source is embedded in G2 could come in the next 5-10 years, when a decoupling between it and the previously outflowing gas might happen after pericentre, due to the increased cross section of the latter. At that point, the hydrodynamical interaction with the accretion flow would act on G2, but not on its central source and the newly emitted material, leading to the decoupling. The luminosity of the outflow material after pericentre can strongly depend on processes that cannot be too reliably captured by the present simulations, particularly during and right after the pericentre passage (see discussion is Sec. \ref{seclum}). The gas lost by the source before the pericentre passage in our simulations \citep[and in those by][]{Schartmann12, Guillochon14, Schartmann15} is decelerated by the hydrodynamical drag of the external accretion flow. At the same time, it is heating up, partially due to the mixing with the outer hot material, eventually leading to a substantial drop of its luminosity. Unfortunately, the mixing in the present simulations is mainly numerical. For this reason, no strong quantitative statement can currently be made, e.g., on the luminosity of old and new material and on the exact time of their decoupling. For our model we can, however, predict a non-symmetric behaviour of the gas, around the pericentre position, along its orbit (as opposed to what is expected for a purely ballistic diffuse cloud), and a ``rebirth'' of G2.

\section{Summary}\label{secsum}

In this work we presented 3D AMR simulations for a ``compact source'' scenario for G2, for which its gas component is produced by an outflow from a central source. Such a study is a natural follow-up of the study by \citet{Ballone13}, performed by means of 2D higher resolution simulations.

We can draw the following strong conclusions:

\begin{enumerate}
\item Relatively massive ($\dot{M}_\mathrm{w}=5\times 10^{-7} \;M_{\odot} \;\mathrm{yr^{-1}}$) and slow ($50 \;\mathrm{km \;s^{-1}}$), compared to main-sequence stars, outflows are needed to reproduce the emission properties of G2; furthermore, the central source must be a low mass star, due to observational constraints. This suggests that a possible source for G2 is a young stellar object, possibly a T Tauri star.
\item The appearance of such an outflow in the PV diagrams is strongly dependent on how much of its unperturbed region is actually emitting; if the material at distances smaller than roughly 100 AU from the source dominates the emission, G2 would always look too compact - both in size and in velocity - compared to the observations.

\item A reasonable comparison to the current SINFONI observations can be obtained both by the diffuse cloud simulations in \citet{Schartmann15} and by the present ones. However, we might be able to understand whether G2 is generated by a source or if it is a simple gas-dust diffuse cloud in the next 5-10 years. For the case of a compact source, we should then be able to observe a decoupling between the dust and gas components and a new and ``fresh'' G2 should reform around the dusty one, later on.
\end{enumerate}

Studying the ``compact source'' model presents more complications, compared to the ``diffuse cloud'' one. Still, the present can reproduce the Br$\gamma$ observations and it has the advantage of being able to explain the simultaneous compactness of G2's dust component and extendedness of its gaseous one.

\section*{Acknowledgements}

This project was supported by the Deutsche Forschungsgemeinschaft (DFG) priority program 1573 ``Physics of the Interstellar Medium'' and the DFG Cluster of Excellence ``Origin and Structure of the Universe''. Computer resources for this project have been provided by the Leibniz Supercomputing Center under grants: h0075, pr86re. Alessandro Ballone would like to thank Michela Mapelli, Andrea Gatto, James Guillochon, Jorge Cuadra, Diego Calder\'on and his PGN colleagues for useful discussions. Most of the simulation post-processing was carried out with the yt toolkit \citep{Turk11}




\bibliographystyle{mnras}
\bibliography{mylit} 








\bsp	
\label{lastpage}
\end{document}